\documentclass[journal,twoside]{IEEEtran}
\usepackage{cite}
\usepackage{xcolor}
\usepackage{amsmath,amssymb,amsfonts}

\DeclareMathOperator*{\argmin}{arg\,min}
\usepackage{subfig}
\usepackage{algorithmic}
\usepackage{bm}
\usepackage{makecell}
\usepackage{multirow}
\usepackage{graphicx}
\usepackage{textcomp}
\usepackage{wrapfig}
\usepackage{balance}
\def\BibTeX{{\rm B\kern-.05em{\sc i\kern-.025em b}\kern-.08em
    T\kern-.1667em\lower.7ex\hbox{E}\kern-.125emX}}

\begin{document}
\title{Robust Deep Compressive Sensing with Recurrent-Residual Structural Constraints}
\author{Jun Niu
\thanks{This is a pre-print.}
\thanks{Jun Niu is now with Amazon (e-mail: j.niu1990@gmail.com). This work is done at Department of Electrical and Computer Engineering, Duke University, Durham, North Carolina, 27708, USA, before him joining Amazon.}
}

\maketitle

\begin{abstract}
Existing deep compressive sensing (CS) methods either ignore adaptive online optimization or depend on costly iterative optimizer during reconstruction. This work explores a novel image CS framework with recurrent-residual structural constraint, termed as R\textsuperscript{2}CS-NET. The R\textsuperscript{2}CS-NET first progressively optimizes the acquired samplings through a novel recurrent neural network. The cascaded residual convolutional network then fully reconstructs the image from optimized latent representation. As the first deep CS framework efficiently bridging adaptive online optimization, the R\textsuperscript{2}CS-NET integrates the robustness of online optimization with the efficiency and nonlinear capacity of deep learning methods. Signal correlation has been addressed through the network architecture. The adaptive sensing nature further makes it an ideal candidate for color image CS via leveraging channel correlation. Numerical experiments verify the proposed recurrent latent optimization design not only fulfills the adaptation motivation, but also outperforms classic long short-term memory (LSTM) architecture in the same scenario. The overall framework demonstrates hardware implementation feasibility, with leading robustness and generalization capability among existing deep CS benchmarks.
\end{abstract}

\begin{IEEEkeywords}
Compressive sensing, structural constraint, adaptive online optimization, recurrent neural network.
\end{IEEEkeywords}

\IEEEpeerreviewmaketitle

\section{Introduction}
\label{sec:introduction}
Information compression and reconstruction stands at the center of communication and signal processing. Compressive sensing (CS) has been an emerging strategy to recover original signals with considerably fewer measurements at sub-Nyquist sampling rate \cite{4286571,4472240}. Different from auto-encoders, where both encoder and decoder can be heavily engineered, conventional CS seeks to alleviate the sampling complexity while reconstructing signals from low-dimensional measurements via computationally intensive optimization. Its sampling efficiency accommodates limited sensor-side computational resources and transmission bandwidth. Compared with deploying complex encoder at the frontend, CS serves as a more economic and realistic option. The online optimization further offers extra robustness to amend measurement noise during reconstruction. Benefiting from its flexible and efficient architecture, CS has been successfully applied in scenarios where measurements are expensive and noisy. Typical applications include magnetic resonance imaging (MRI) and high resolution imaging systems \cite{Ye2019,pmlr-v97-wu19d}.

Classic CS theory assumes the signal to be sparse or can be expressed as a sparse linear combination of known basis. This strong requirement hinders CS's practical application as such basis is implicit in general. Recently, Bora et al. introduces deep learning based structural constraint to relax this requirement \cite{pmlr-v70-bora17a}. A \textit{pre-trained} deep neural network (NN) is proposed as the structural constraint in the place of sparsity, which suffices a generalized set-restricted eigenvalue condition. With deep structural constraint and successful training, low reconstruction error for dense signals can still be achieved with high probability. The reconstruction, however, heavily relies on classic iterative optimization. Its slow convergence remains a critical challenge in practice. In addition, the sampling relies on random measurement matrices, which is sub-optimal for highly structured signals such as natural images.

Inspired by deep NN's powerful ability in signal representation, several deep CS methods are proposed to enhance the performance of both sampling and reconstruction. For example, Wu S. et al. \cite{pmlr-v97-wu19b} proposed an unrolled feed-forward NN for learning the measurement matrix, while depending on classic iterative optimization for reconstruction. Kulkarni et al. proposed a non-iterative deep reconstruction model, the ReconNet \cite{8379450}. It achieves high efficiency and robustness with convolutional neural network (CNN), though the architecture does not emphasize domain insights. The DR\textsuperscript{2}-NET \cite{YAO2019483} applies residual learning for reconstruction enhancement. The MS-DCI \cite{9256346} develops an elegant multi-scale sampling framework with cascaded multi-stage reconstruction. The multi-level sampling decomposition, however, potentially increases the sampling hardware complexity or transmission bandwidth. More recent frameworks, such as OPINE-NET \cite{9019857} and AMP-NET \cite{9298950}, unfold iterative optimization algorithms for model based methods into networks. Though efficient and interpretable, the unrolled networks possess reduced robustness and generalization capacity versus the iterative online optimizers. As an alternative, recently Wu Y. et al. \cite{pmlr-v97-wu19d} combines explicit online optimization with learnable deep structural constraint and measurement matrix. This framework largely speeds up the convergence of online optimization. However, it still relies on the classic iterative gradient descent optimizer for adaptive optimization, which is computationally costly and slow. 

This work proposes a novel framework, termed as R\textsuperscript{2}CS-NET, to effectively bridge deep image CS with adaptive online optimization. In particular, a novel recurrent neural network (RNN) is explored as an alternative for unrolled NN during initial reconstruction. It aims to enhance the structural constraint's capacity, such that the overall system operates more robustly and better generalizes in online usage, while avoiding expensive operations like iterative gradient calculation. The R\textsuperscript{2}CS-NET is composed of three stages: measurement, recurrent latent optimization, and residual convolutional reconstruction. The measurement stage assembles a learnable, modified structural random matrix \cite{6041037}. It is the only module deployed at the sensor frontend, and exploits switchable pre-coded masks to adapt the R\textsuperscript{2}CS-NET in different sampling domains without re-train necessity. The recurrent latent optimization attempts to acquire the robustness and generalization capacity of online optimization. This stage is established through a novel recurrent pipeline. It progressively optimizes the latent representation of the acquired samplings to accommodate the downstream reconstruction network. The residual CNN then fully reconstructs the original image in pixel domain. The recurrent latent optimization network and cascaded residual CNN together serve as the structural constraint in the place of sparsity. All components are fully convolutional for concurrency acceleration. The three stages are optimized jointly during training toward optimal reconstruction quality under given sampling rate.

Signal correlations have been addressed through the network architecture. The adaptive sensing nature of  R\textsuperscript{2}CS-NET further makes it an ideal candidate for color image CS. Numerical experiments verify the efficiency and robustness of the proposed design. The R\textsuperscript{2}CS-NET also demonstrates leading performance among recent deep CS benchmarks.

In summary, this work contributes the following.
\begin{itemize}
  \item The R\textsuperscript{2}CS-NET is the first deep CS framework bridges efficient online optimization with unrolled reconstruction network. Expensive operations such as gradient computations are completely avoided during reconstruction.
  \item A novel recurrent-residual architecture is explored to enhance the generalization capacity and robustness of deep structural constraints.
  \item The R\textsuperscript{2}CS-NET intrinsically supports color image CS. The overall system exhibits leading performance among deep CS benchmarks. 
\end{itemize}

\section{Deep Compressive Sensing with Structural Constraints and Online Optimization}
\label{section: cs theory}
Compressive sensing aims to efficiently acquire a compressible signal $\bm{x}$ and then reconstruct the original signal from samplings $\bm{m}$ via finding optimal solutions to the under-determined linear equation system. Exploiting signal's sparsity or incoherence nature, the original signal can be recovered with far fewer samplings required by the Nyquist–Shannon sampling theorem. Formally,
\begin{equation}
\label{eq: cs theory}
\bm{m} = \mathbf{F} \bm{x} + \bm{\eta} ,
\end{equation}
where $\mathbf{F}$ is the measurement matrix and $\bm{\eta}$ denotes the noise at measurement time. Depending on the sparsity of original signal $\bm{x}$, the dimension of measured signal $\bm{m}$ is usually considerably smaller than that of $\bm{x}$.

A necessary and sufficient condition for this problem to be well conditioned is the Restricted Isometry Property (RIP). It states, for a small constant $\delta \in (0, 1)$, the difference between two signals $\bm{x}_1 - \bm{x}_2$ satisfies
\begin{equation}
\begin{split}
\label{eq: RIP loss}
(1 - \delta) || \bm{x}_1 - \bm{x}_2 ||^2 & \leq || \mathbf{F} (\bm{x}_1 - \bm{x}_2) ||_2^2 \\
& \leq (1 + \delta) || \bm{x}_1 - \bm{x}_2 ||_2^2.
\end{split}
\end{equation}
In other words, the measurement matrix $\mathbf{F}$ preserves the distance between original sparse signals with small perturbation \cite{pmlr-v97-wu19d}. The original signal thus can be recovered by solving the optimization problem

\begin{equation}
\label{eq: cs reconstruction}
\hat{\bm{x}} = \argmin_{\bm{x}} || \bm{m} - \mathbf{F} \bm{x} ||_2^2.
\end{equation}

Classic CS theory requires $\bm{x}$ to be a sparse signal or a dense signal that can be expressed as a sparse linear combination of known basis vectors. This enforces a strong restriction. In most practical problems the sparse basis vectors are not explicit to discover, sometimes extremely untrivial to obtain. 

As an alternative, Bora et al. \cite{pmlr-v70-bora17a} propose using a \textit{pre-trained} deep NN as the structural constraint in the place of sparsity
\begin{equation}
\label{eq: structural constraint}
\bm{x} = G (\bm{z}).
\end{equation}
The network $G(\cdot)$ implicitly constrains the compressibility through its architecture, where the latent space tensor $\bm{z}$'s dimension is usually considerably smaller than that of $\bm{x}$. This condition is sufficient to provide a generalized Set-Restricted Eigenvalue Condition (S-REC). It guarantees that low reconstruction error can be achieved with high probability.

Instead of directly solving Eq. (\ref{eq: cs reconstruction}), the reconstruction now first estimates the latent representation $\bm{z}$ by solving the online optimization problem
\begin{equation}
\label{eq: latent space reconstruction}
\hat{\bm{z}} = \argmin_{\bm{z}} || \bm{m} - \mathbf{F} G(\bm{z}) ||_2^2,
\end{equation}
then reconstructs the original signal through $\hat{\bm{x}} = G(\hat{\bm{z}})$.

The structural constrained CS framework connects deep learning with classic CS theory. It provides a theoretical foundation to tackle arbitrary dense, compressible signals in data-driven manner. A few deep CS methods have been proposed thereafter. For example, the reconstruction can be accomplished through either classic optimizers or unrolled NN designs; the structural constraint $G$ can be learnt together with the measurement matrix to achieve better reconstruction quality \cite{pmlr-v97-wu19d,9019857,Shi_2019_CVPR,9159912,8953841}. However, whenever adaptive online optimization is leveraged during the reconstruction, the gradient-based optimization of Eq. (\ref{eq: latent space reconstruction}) has to be addressed in every  iteration. As the structural constraint typically consists of heavy-weight NN, repeated evaluation of  $\frac{\partial G(\bm{z})}{\partial \bm{z}}$ along the iterative optimization turns out to be extremely expensive. The computational cost further increases with more complex architecture designs. The unrolled NN reconstruction sidesteps the cost concerns. However, ignoring the adaptive online optimization potentially leads to sub-optimal solution with non-perfect $G(\cdot)$, impairing the system's robustness under the inevitable measurement noise in practice. An efficient, robust deep CS solution remains at the center of the problem.

\section{R\textsuperscript{2}CS-NET: A Novel Image Compressive Sensing Framework}

This work introduces a novel framework to efficiently bridge deep CS with adaptive online optimization. The proposed R\textsuperscript{2}CS-NET consists of three main stages: measurement, recurrent latent optimization, and residual convolutional reconstruction. The measurement stage is engineered with simplicity and hardware implementation feasibility. With acquired samplings, an RNN first progressively adapts the samplings' latent representation to the downstream reconstruction network. The residual CNN then fully reconstructs the original image. The recurrent latent optimization and the residual CNN together serve as the recurrent-residual structural constraint. It delivers higher generalization capacity and robustness than unrolled networks at moderate latency.

\begin{figure*}[ht]
\begin{center}
\centerline{\includegraphics[width=1.0\textwidth]{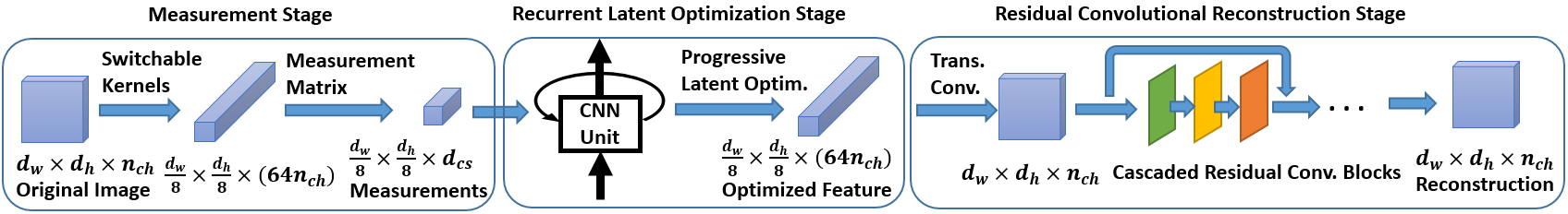}}
\end{center}
\caption{Systematic sketch of the R\textsuperscript{2}CS-NET. The system consists of three stages: measurement, adaptive recurrent optimization, and residual convolutional reconstruction. Only the measurement stage needs to be deployed with frontend sensor hardware.}
\label{fig: systematic sketch}
\end{figure*}

Each stage has its specific motivation. However, the three objectives are not mutually independent. All stages can be cascaded into an end-to-end system during training.  A system sketch is illustrated in Fig. \ref{fig: systematic sketch}. The network architecture exploits signal correlation for optimal sampling efficiency.

\section{Measurement}
The measurement stage is the only module deployed with sensors. One primary difference between CS and auto-encoders is the emphasis on efficiency and simplicity of data acquisition. The measurement stage is therefore engineered with hardware compatibility in mind. It consists of two fully convolutional layers, one with switchable pre-coded kernels adapting the R\textsuperscript{2}CS-NET from various data acquisition spaces to the preferred sampling space; the other operating as the standard measurement matrix. The measurement matrix is the only component to be learnt from data. 

\begin{figure}[ht]
\begin{center}
\centerline{\includegraphics[height=0.35\columnwidth]{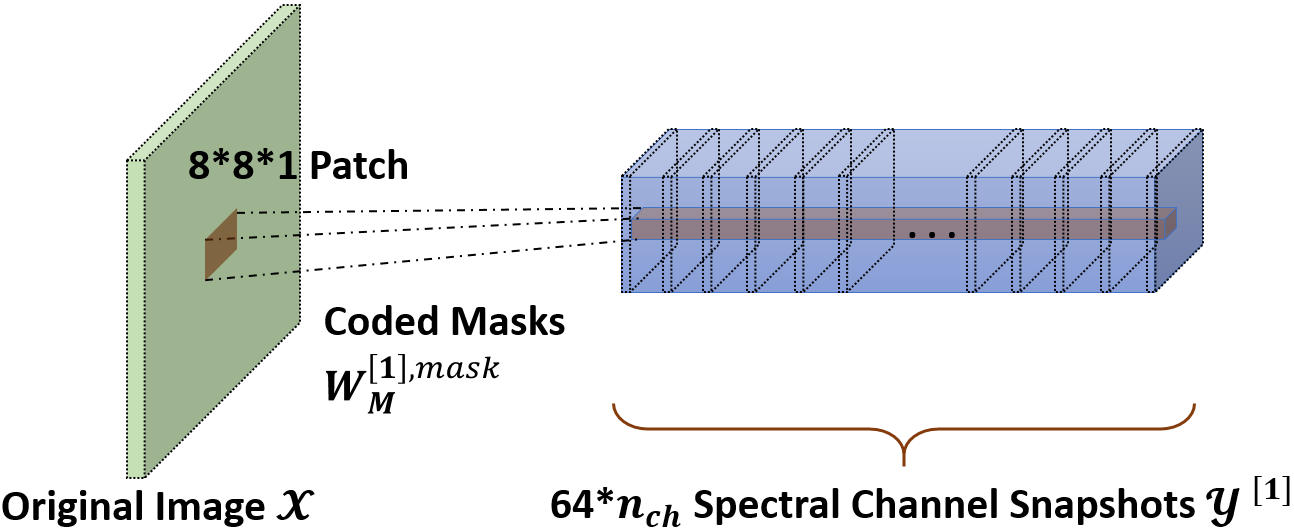}}
\end{center}
\caption{Mapping input images from acquisition domain to k-space. The $\mathbf{W}_{M}^{[1], mask}$ kernel is a set of $64 \times n_{ch}$ coded masks.}
\label{fig: measurement dct mapping}
\end{figure}

The first layer's operation can be written as
\begin{equation}
\bm{\mathcal{Y}}^{[1]} = \mathbf{W}_{M}^{[1], mask} * \bm{\mathcal{X}},
\end{equation}
where $*$ denotes the 2D convolution operator; $\bm{\mathcal{X}}$ is the image in acquisition domain; $\mathbf{W}_{M}^{[1], mask}$ is the adaptive coded mask. Convolution stride is chosen as $[N_1, N_2]$ along width and height dimension, respectively; $N_1$ and $N_2$ denotes the width and height of each sampling block. This layer has no trainable parameters. The parallelism nature of convolution operations further guarantees concurrency efficiency. Figure \ref{fig: measurement dct mapping} depicts the mapping mechanism. 

Without loss of generality, we consider pixel domain sampling as the default setting. The front layer maps pixel value into k-space to exploit natural images' spectral domain sparsity. Inspired by this operation's discrete cosine transform (DCT) nature, the following set of coded masks are plugged in place of convolutional kernels.
\begin{equation}
\begin{split}
\label{eq: DCT coded mask}
& \mathbf{W}^{DCT}_{i,j, k, k + k_1 \cdot N_{ch} + k_2 \cdot N_1 N_{ch}}  =  \sqrt{\frac{2}{N_1}} \sqrt{\frac{2}{N_2}} \alpha (k_1) \alpha(k_2) \cdot \\
& \qquad\qquad  \cos [\frac{(2n_1 + 1) \pi k_1)}{2 N_1}] \cos [\frac{(2n_2 + 1) \pi k_2}{2 N_2}] , \\
& \qquad i = 0, 1, ..., N_1 - 1, j = 0, 1, ..., N_2 - 1, \\
& \qquad k = 0, ..., N_{ch} - 1,
\end{split}
\end{equation}
where $N_{ch}$ denotes the colorspace channel. The measurement then multi-scales on frequency bands of input images. This step largely relieves the complexity expectation of the followed operation, which now can solely focus on linear measurement without further domain transformation. When directly acquiring data in k-space, one can simply replace the above with a set of binary coded masks \cite{9206641}. The entire R\textsuperscript{2}CS-NET thereafter operates in the new sampling space without re-train necessity. Other acquisition spaces can be adapted accordingly.

The second convolutional layer samples the original image block by block, across image's channels
\begin{equation}
\label{eq: cs measurement}
\bm{\mathcal{M}} = \mathbf{W}_{M}^{[2]} * \bm{\mathcal{Y}}^{[1]},
\end{equation}
where $\mathbf{W}_{M}^{[2]}$ is the learnable CS measurement matrix; $\bm{\mathcal{M}}$ is the tensor of measured samplings. The convolution kernel is set to be of size $[1, 1]$ with stride 1 for block-wise measurement. Each tensor slice in $\bm{\mathcal{M}}$ with unit width and unit height thus represents the measurements in one sampling block.

One observation is the above two layers assemble a modified, learnable structural random matrix (SRM) \cite{6041037}. The SRM is defined as a product of matrices $\bm{\Phi} = \bf{D}\bf{T}\bf{R}$, where $\bf{D}$ is the subsampling matrix; $\bf{T}$ is the transform kernel; $\bf{R}$ is a randomizer. $\mathbf{W}_{M}^{[2]}$ and $\mathbf{W}_{M}^{[1], mask}$ perform in the place of $\bf{D}$ and $\bf{T}$, respectively.

The measurement stage can be feasibly integrated on hardware with coded aperture cameras \cite{Wagadarikar:08} or CS-CMOS cameras \cite{5548164}. For example, the switchable coded mask $\mathbf{W}_{M}^{[1], mask}$ and the learnt measurement matrix $\mathbf{W}_{M}^{[2]}$ can be implemented as high resolution coded apertures modulating the optical field from the scene, with a low resolution focal plane aperture instated for sampling data acquisition. Only the coded aperture needs to be switched when operating pre-trained R\textsuperscript{2}CS-NET in new sampling space. The computationally inexpensive coded mask and measurement matrix can also be realized in digital domain as on CS-CMOS cameras, whose on-chip processor is strictly restricted by limited power and area on sensors.

\section{Recurrent-Residual Structural Constraint}

The recurrent-residual structural constraint aims to bridge the advantages of adaptive online optimization with deep CS, while avoiding the intensive costs of classic iterative optimizers. A two-phase layout is proposed to this end. The recurrent latent optimization stage resembles the gradient descent optimization through a novel RNN architecture. It preliminarily updates the samplings in latent space. The cascaded residual CNN fully reconstructs the image thereafter. Compared with the unrolled NN methods, the recurrent-residual structural constraint better generalizes on input patterns. Though not citing particular noise prior, it expands the robustness under measurement noise through architecture design with successful data-driven training.
\subsection{Recurrent Latent Representation Optimization}
\label{section: recurrent latent space optimization}

The latent representation can be generally of arbitrary shape. Aiming at preliminarily reconstructing toward the original image, here $\bm{z}$ is designed with the same dimension as $\bm{\mathcal{Y}}^{[1]}$. The gradient descent latent optimization is thus formulated as
\begin{equation}
\begin{split}
\label{eq: naive latent optimization}
\hat{\bm{\mathcal{Z}}}_{i+1} = & \hat{\bm{\mathcal{Z}}}_i - \alpha \frac{\partial}{\partial \hat{\bm{\mathcal{Z}}}_i} || F (\hat{\bm{\mathcal{Z}}}_i) - \bm{\mathcal{M}}||_2^2,
\end{split}
\end{equation}
where the subscript $i$ denotes the iteration step; $\hat{\bm{\mathcal{Z}}}_{i+1}$ is progressively updated through the incremental adjustment to $\hat{\bm{\mathcal{Z}}}_i$.

Our key observation is the stochastic gradient descent update of Eq. (\ref{eq: naive latent optimization}) resembles the cell state update in classic RNNs \cite{10.5555/2969239.2969329},
\begin{equation}
\mathcal{C}_t = f_t \odot \mathcal{C}_{t-1} + i_t \odot \tilde{\mathcal{C}}_{t},
\end{equation}
where the subscript $t$ denotes the recurrent step; $-i_t$ operates as the learning rate and the new candidate state $\tilde{\mathcal{C}}_{t}$ corresponds to the incremental adjustment; $f_t$ operates as the reset coefficient for the previous state; $\odot$ denotes the Hadamard product.  

This encourages us to design a recurrent unit assembling similar online latent optimization. Each iteration in Eq. (\ref{eq: naive latent optimization}) explicitly depends on the previous output $\hat{\bm{\mathcal{Z}}}_i$ and the original measurements $\bm{\mathcal{M}}$. This is robust in clean sampling environment. However, with the presence of measurement noise, the optimization deviates from the reconstruction goal due to the noisy $\bm{\mathcal{M}}$. We consider a similar RNN. At each step, the recurrent unit consumes $\hat{\bm{\mathcal{Z}}}_i$ and $F (\hat{\bm{\mathcal{Z}}}_i) - \bm{\mathcal{M}}$ as inputs. In noisy sampling environment both $F (\hat{\bm{\mathcal{Z}}}_i)$ and $\bm{\mathcal{M}}$ are perturbed. Their absolute difference along the iterations converges towards a non-zero random variable. This residual potentially indicates environmental attributes such as noise variance, and assists the RNN amending the latent representation.

The second intuition is the wavelet decomposition nature of measurement residuals. The linear measurement operation maintains the residuals in the same space along the optimization steps. This is comparable to a wavelet sequence. One transform that successfully updates one packet is highly likely to successfully update the next packet in the same space.

A novel RNN is proposed based on the above analysis to progressively optimize the samplings' latent representation through measurement residuals. At each step, the recurrent unit operates upon the residual between the original samplings and measurements of previous step's output. The residual eventually converges to zero in noise-free environment. In noisy measurement environment, it highlights the noise attributes for further denoising amendment. This architecture together with successful data-driven training serves as the prior that enhances the system's noise robustness. One practical concern, however, is the measurement residuals vary in magnitudes over the recurrence. The recurrent unit is expected to possess consistent performance over wide bandwidth. To tackle this challenge, a scaling gate network renders the residual input with appropriate gain before generating the new candidate. $i_t$ and $f_t$ are both designed as CNNs to further enhance the recurrent unit's capacity.

Summarily, at the initial step, the input $\bm{\mathcal{I}_0}$ is initialized with the measurement sampling $\bm{\mathcal{M}}$. At each following step $i$, the recurrent unit updates the latent representation $\bm{\mathcal{Z}_i}$ as
\begin{equation}
\label{eq: recurrent unit}
\begin{split}
\bm{\mathcal{I}_i} &= S(F(\bm{\mathcal{Z}_{i-1}}) - \bm{\mathcal{I}_{0}}), \\
\bm{\mathcal{H}_i} &= C(\bm{\mathcal{I}_i}), \\
\bm{\mathcal{Z}_i} &=  R(\bm{\mathcal{Z}_{i-1}}, \bm{\mathcal{H}_i}) \odot \bm{\mathcal{Z}_{i-1}} +  U(\bm{\mathcal{Z}_{i-1}}, \bm{\mathcal{H}_i}) \odot \bm{\mathcal{H}_i}, \\
& i = 1, 2, ..., T.
\end{split}
\end{equation}
Here $S(\cdot), C(\cdot), R(\cdot), U(\cdot)$ are all CNNs instead of coefficients; $F(\cdot)$ is the measurement operation following Eq. (\ref{eq: cs measurement}). The scaling gate $S(\cdot)$ adjusts the residual input with appropriate gain at each step. The candidate gate $C(\cdot)$ generates a new candidate from scaled input. The latent representation is then updated as a trade-off between previous result and new candidate, where the reset gate $R(\cdot)$ and the update gate $U(\cdot)$ calculate the trade-off coefficients, respectively.

\begin{figure}[ht]
\begin{center}
\subfloat[][]{\includegraphics[height=0.45\columnwidth]{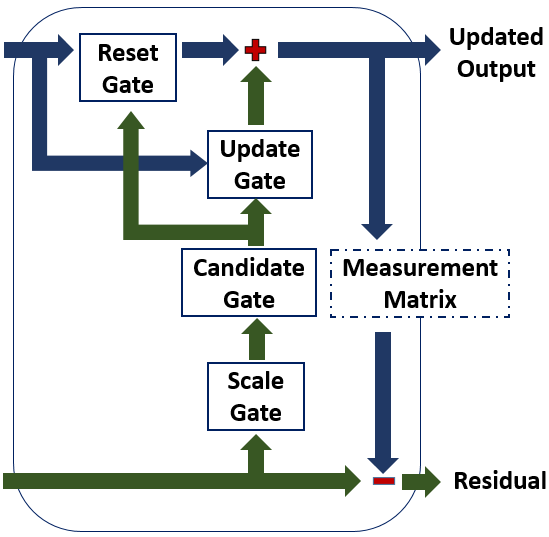}} \hfill 
\subfloat[][]{\includegraphics[height=0.42\columnwidth]{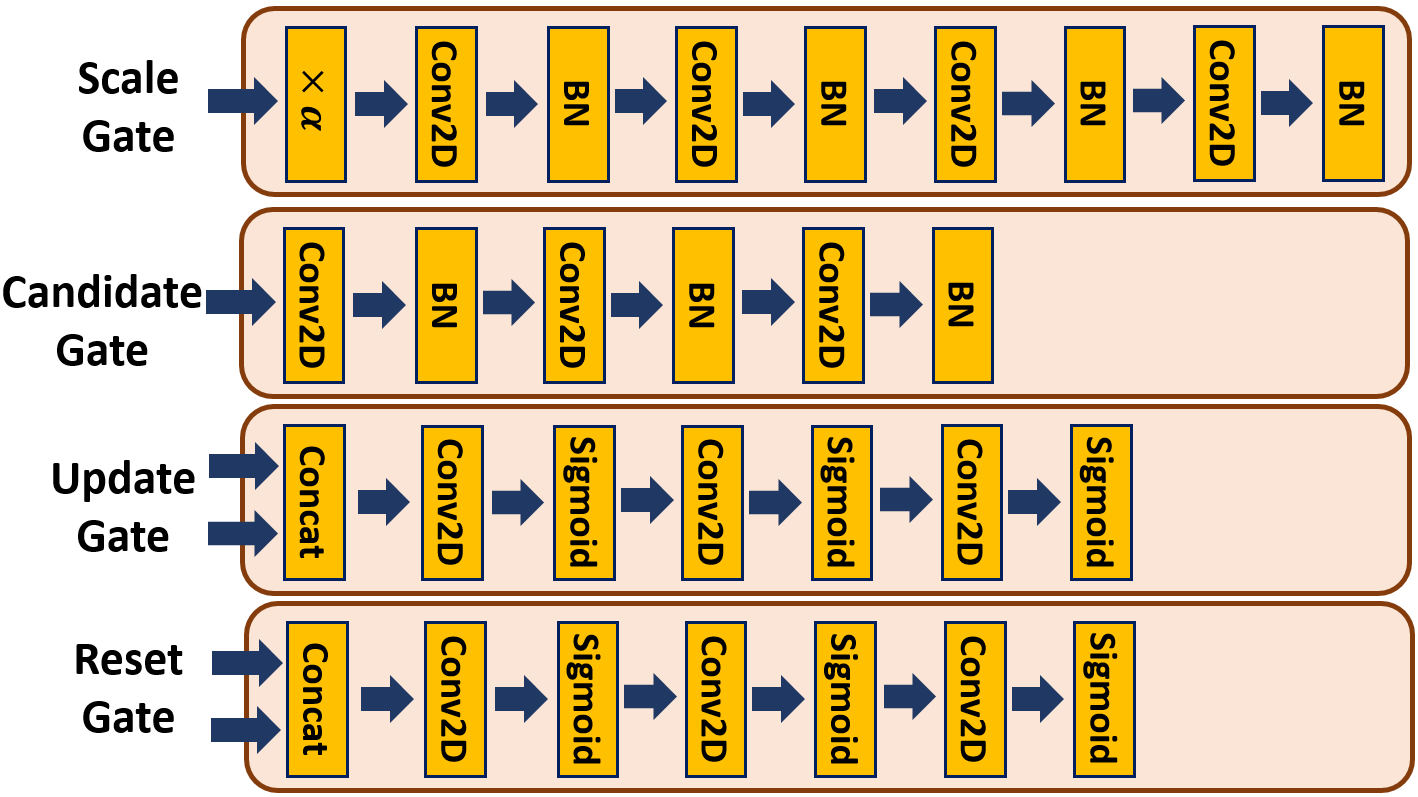}}
\end{center}
\caption[caption]{Recurrent unit architectures for latent optimization: (a) the overall sketch; (b) network designs for each inner gate.} 
\label{fig: recurrent unit}
\end{figure}

Conventional LSTM and GRU cells use dense connection architecture. This contradicts with the uncorrelated nature of samplings measured from different image blocks, simultaneously resulting in numerous unnecessary parameters. As an alternative, $S(\cdot), C(\cdot), R(\cdot), U(\cdot)$ consist of 2D convolution layers with kernel size $[1, 1]$ and stride 1. Operating on the latent representation sampling block by sampling block, it successfully avoids block-wise coupling. A batch normalization layer is further added at the recurrent input and output to enhance nonlinear processing capability.

\begin{figure}[ht]
\begin{center}
\centerline{\includegraphics[width=0.5\textwidth]{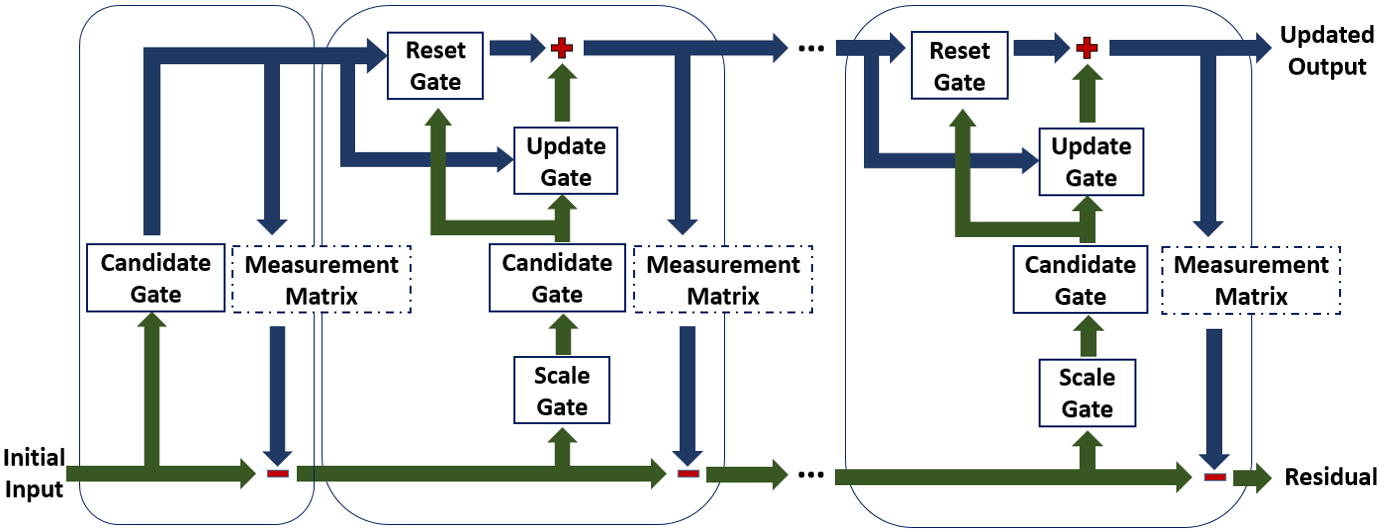}}
\end{center}
\caption{Computational pipeline for recurrent latent optimization. The initial step evaluates on the measurement samplings. The following steps first scale the residual input and then update latent representation with trade-off.}
\label{fig: recurrent pipeline}
\end{figure}

Figure \ref{fig: recurrent unit} - \ref{fig: recurrent pipeline} depicts the proposed recurrent unit architecture and computation pipeline, respectively. At the initial step the candidate gate directly operates on the measured samplings $\bm{\mathcal{M}}$. Latent representation then receives progressive updates from previous step's output and measurement residual. All network parameters are shared across the recurrent steps.

\subsection{Residual Convolutional Reconstruction}

This stage reconstructs the original image from the optimized latent representation $\bm{\mathcal{Z}}$. Formally, the residual CNN $G(\cdot)$ aims to reconstruct an image $\hat{\bm{\mathcal{X}}} =  G(\bm{\mathcal{Z}})$  such that $\hat{\bm{\mathcal{X}}}$ is as close to the original image $\bm{\mathcal{X}}$ as possible. Sharpness enhancement and artifacts suppression need to be accomplished toward this goal.

Successful reconstruction relies on comprehending the intrinsic correlation of $\bm{\mathcal{Z}}$'s components. Here we first enrich the latent representation through cascaded convolutions
\begin{equation}
\begin{split}
\label{eq: conv2d with bn}
G^{[l]} (\bm{\mathcal{Z}}) = BN(\mathbf{W}_{G}^{[l]} * G^{[l - 1]} (\bm{\mathcal{Z}}) + \mathbf{B}_{G}^{[l]}), l = 1, 2
\end{split}
\end{equation}
where $BN(\cdot)$ denotes a batch normalization layer; superscript $[l]$ denotes the layer index. $\mathbf{W}_{G}^{[1]}$ and $\mathbf{W}_{G}^{[2]}$ contain $N_1 \times N_2 \times N_{ch}$ channels. The block-wise latent optimization addresses the inter-channel correlation while maintaining the relative independency across adjacent sampling blocks. As a result, both kernels are correspondingly designed as size $[1, 1]$  with stride 1.

A transposed convolution layer then maps $G^{[2]} (\bm{\mathcal{Z}})$ to the original image's shape
\begin{equation}
\begin{split}
\label{eq: deconv}
\tilde{G}^{[3]}(\bm{\mathcal{H}}) = {\mathbf{C}_{G}^{[3]} }^{T} \cdot \tilde{G}^{[2]}(\bm{\mathcal{Z}}) + \mathbf{B}_{G}^{[3]} ,
\end{split}
\end{equation}
where $\cdot$ denotes the matrix multiplication; $\mathbf{C}_{G}^{[3]}$ is the matrix operator of the transposed convolution; $\tilde{G}^{[2]} (\cdot)$ and $\tilde{G}^{[3]} (\cdot)$ denote the row-major order vector representation of the tensor outputs of $G^{[2]} (\cdot)$ and $G^{[3]} (\cdot)$, respectively. After this operation, the latent representation is transformed to be in same shape as $\bm{\mathcal{X}}$, while maintaining the same number of entries.

The initial macroblocking artifacts suppression is achieved with three cascading convolution layers with the same operations described by Eq. (\ref{eq: conv2d with bn}). The transposed convolution maps entries along tensor channels to width and height dimensions. As a result, the convolution kernels of shape $[11, 11]$, $[7, 7] $ and $[1, 1]$ are used to exploit the intra-channel spatial correlation of $G^{[3]}(\bm{\mathcal{Z}})$. To guarantee the initial decoding results falling into the non-negative pixel value range, a rectifier linear unit (ReLU) function activates the output of $G^{[6]}$.

With macroblocking artifacts mitigated, detail enhancement remains the next objective towards high quality reconstruction. Here we apply cascaded residual CNNs for image sharpening. Operations in each residual block can be formulated as 
\begin{equation}
\begin{split}
\label{eq: residual CNN}
& G^{[l]} (\bm{\mathcal{H}}) = \\
   & \begin{cases}
      \makecell{ \mathbf{W}_{G}^{[l]} * G^{[l - 1]} (\bm{\mathcal{H}})  + \mathbf{B}_{G}^{[l]} +  G^{[l - 3]} (\bm{\mathcal{H}})  ,} & \makecell{  l = 9, 12, \\ 15, 18, 21 } \\
      \makecell{ ReLU(BN(\mathbf{W}_{G}^{[l]} * G^{[l - 1]} (\bm{\mathcal{H}}) + \mathbf{B}_{G}^{[l]})), } & \makecell{ l = 7, 8, ..., 20.}
    \end{cases}
\end{split}
\end{equation}

This architecture conceptually decomposes the image into wavelet series. The convolution layers inside each residual block focuses on sharpening high frequency details. After merging with the low frequency image snapshots, features of the complete image are delivered to the next cascaded residual block for further enhancement.

Passing through five cascaded residual convolution blocks, the final reconstruction is activated with a ReLU function:
\begin{equation}
\hat{\bm{\mathcal{X}}} = ReLU( G^{[21]} (\bm{\mathcal{H}}) ).
\end{equation}
It enforces pixel values into the non-negative definition range, and enhances the model's nonlinear capacity in the meantime.

\section{Learning Measurement Matrix and Structural Constraint}
\label{section: training}

Successful learning fulfills each of the R\textsuperscript{2}CS-NET's three stages its particular role. An effective measurement stage enforces the RIP condition while collaborating with the reconstruction stages for optimal sampling efficiency. This objective can be described through the measurement loss \cite{pmlr-v97-wu19d}
\begin{equation}
\label{eq: measurement loss}
L_{M} = \frac{1}{N} \sum_{i=1}^{N} (||F(\hat{\bm{\mathcal{X}}}_i - \bm{\mathcal{X}}_i) ||_2^2 - ||\hat{\bm{\mathcal{X}}}_i - \bm{\mathcal{X}}_i||_2^2 )^2 ,
\end{equation}
where $N$ is the number of training images.

Meanwhile, the recurrent latent optimization stage adapts the samplings to accommodate the downstream reconstruction.  Given a \textit{trained} downstream network, this objective is outlined by the loss function
\begin{equation}
\label{eq: rnn loss}
L_{R} = \frac{1}{N} \sum_{i=1}^{N} || F \cdot G(\hat{\bm{\mathcal{Z}}}_i)  - F(\bm{\mathcal{X}}_i) ||_2^2,
\end{equation}
where $\hat{\bm{\mathcal{Z}}}_i$ denotes the optimized latent representation of samplings; $G(\hat{\bm{\mathcal{Z}}}_i)$ reconstructs $\hat{\bm{\mathcal{X}}}_i$.

The residual convolutional stage minimizes the discrepancy between final reconstructions and the ground truth.  One common formulation is the mean squared error (MSE) loss
\begin{equation}
\label{eq: mse loss}
L_{MSE} = \frac{1}{N} \sum_{i=1}^{N} ||\hat{\bm{\mathcal{X}}}_i - \bm{\mathcal{X}}_i ||_2^2 .
\end{equation}
It equivalently optimizes the reconstructed images' peak signal-to-noise ratio (PNSR) \cite{7115171,9206641}. However, according to specific engineering applications, the loss function can be easily revised to address other perceptual quality metrics \cite{Johnson2016PerceptualLF}.

Each stage can be trained separately toward its goal or iteratively updated in interaction. However, despite the distinct motivations, the three objectives are not mutually independent. $L_{R}$ and $L_{M}$ simultaneously reach their minimum with optimized discrepancy between final reconstruction $\hat{\bm{\mathcal{X}}}$ and the ground truth $\bm{\mathcal{X}}$. Cascading as an end-to-end system during training, all stages can be jointly trained under $L_{MSE}$. This is applied as a preferred training strategy realizing it not only avoids potential slow convergence, but also enforces the optimal sampling - reconstruction of the overall system.

\section{Experiments}

This section provides intensive experiments to validate the efficiency and performance of the proposed R\textsuperscript{2}CS-NET framework. Without loss of generality, we set the recurrent optimization length as $T = 20$ and sampling block size as $8 \times 8 \times N_{ch}$. These hyper-parameters are empirically selected with accommodating the limited computational resources. Though not optimized through intensive hyper-parameter search, the R\textsuperscript{2}CS-NET shows leading performance among existing deep CS benchmarks. The advantage of the proposed recurrent-residual constraint is demonstrated through its enhancement to reconstruction quality and noise robustness under various measurement noise levels. Comparison with classic online optimization methods demonstrates the R\textsuperscript{2}CS-NET achieves comparable sensing fidelity at considerably higher efficiency.

\subsection{Training}
\label{section: training}

The training and validation datasets are cropped from the DIV2K dataset \cite{Timofte_2018_CVPR_Workshops}, which consists of 1,000 high quality color images of 2K resolution. Without loss of generality, we set the dimension of training images as $128 \times 128 \times 3$, where the last dimension denotes the images' RGB channels in pixel domain. The R\textsuperscript{2}CS-NET framework, however, is insensitive to the image size. The entire training dataset contains 1,098,400 images at $128 \times 128$ resolution and the validation dataset has 1000 images at the same resolution. The training, validation and testing datasets throughout the discussion share no overlap.

Assembled as an end-to-end system, all modules are updated interactively by minimizing Eq. (\ref{eq: mse loss}) with Adam optimizer. The recurrent-residual structural constraint enhances the generalization capacity in architecture. This potential needs to be further attained through appropriate data-driven training. An environmental noise is thus engineered to instill the robustness under practical application. Throughout the training process, a white Gaussian noise with standard deviation $\sigma = 0.1$ is added to all measurement operations during sampling and recurrent latent optimization. This perturbation is empirically chosen to the extent that no noticeable reconstruction deterioration is observed in noise-free environment. Once training completes, the R\textsuperscript{2}CS-NET can reconstruct images at arbitrary resolution and operate under arbitrary noise level.

\subsection{Validating the Recurrent Latent Optimization}

\begin{figure}[ht]
\begin{center}
\captionsetup[subfigure]{justification=centering}
\subfloat[][$\bm{\mathcal{Z}_5}$]{\includegraphics[height=0.32\columnwidth]{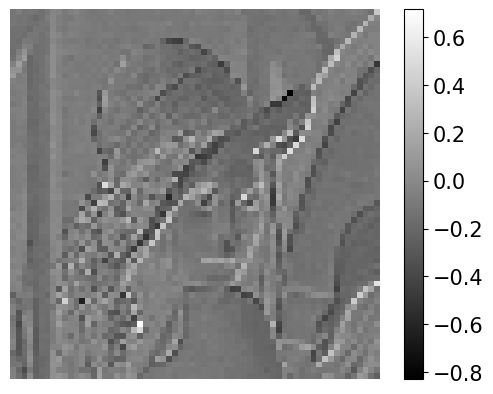}} \hspace{5mm} 
\subfloat[][$\bm{\mathcal{Z}_{10}}$]{\includegraphics[height=0.32\columnwidth]{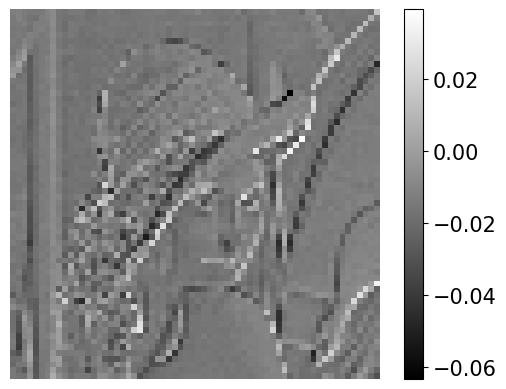}}  
\\[-2ex]
\noindent
\subfloat[][$\bm{\mathcal{Z}_{15}}$]{\includegraphics[height=0.32\columnwidth]{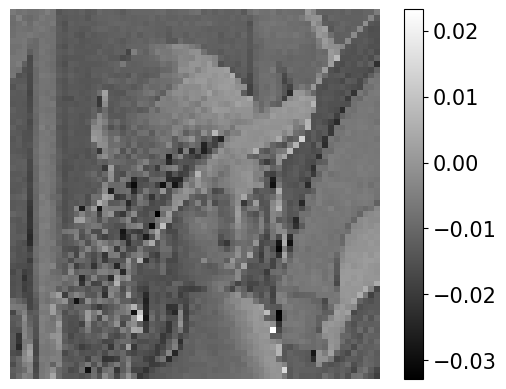}} \hspace{5mm}
\subfloat[][$\bm{\mathcal{Z}_{20}}$]{\includegraphics[height=0.32\columnwidth]{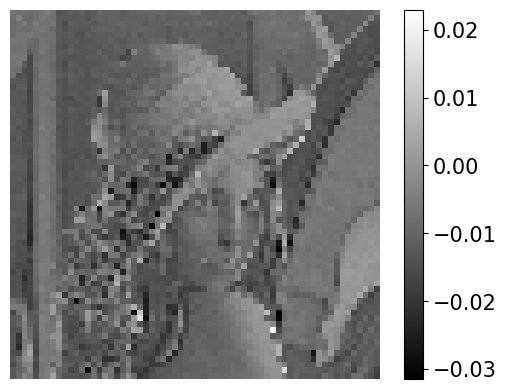}} 
\end{center}
\caption[caption]{Optimized latent representation $\bm{\mathcal{Z}_i}$ at recurrent step $i = 5, 10, 15, 20$, illustrating the tensor's first channel.} 
\label{fig: latent steps}
\end{figure}

\begin{figure}[ht]
\begin{center}
\captionsetup[subfigure]{justification=centering}
\subfloat[][$\hat{\bm{\mathcal{X}}}$ with $i = 5$, PSNR = 4.76, SSIM = 0.070.]{\includegraphics[width=0.35\columnwidth]{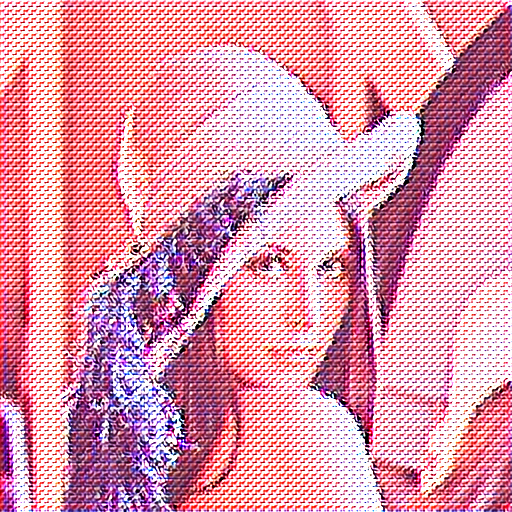}} \hspace{5mm} 
\subfloat[][$\hat{\bm{\mathcal{X}}}$ with $i = 10$, PSNR = 9.96, SSIM = 0.278.]{\includegraphics[width=0.35\columnwidth]{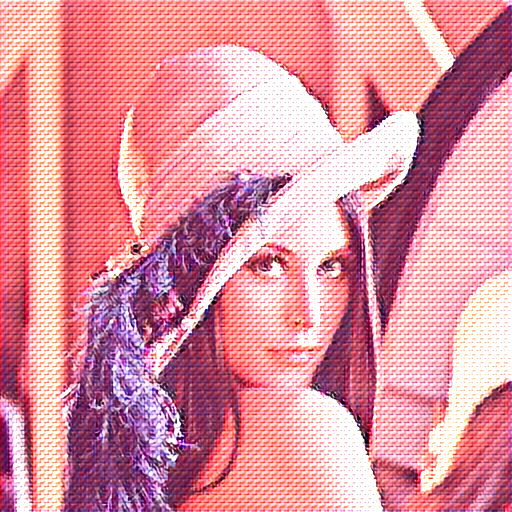}}  
\\[-2ex]
\noindent
\subfloat[][$\hat{\bm{\mathcal{X}}}$ with $i = 15$, PSNR = 17.41, SSIM = 0.720.]{\includegraphics[width=0.35\columnwidth]{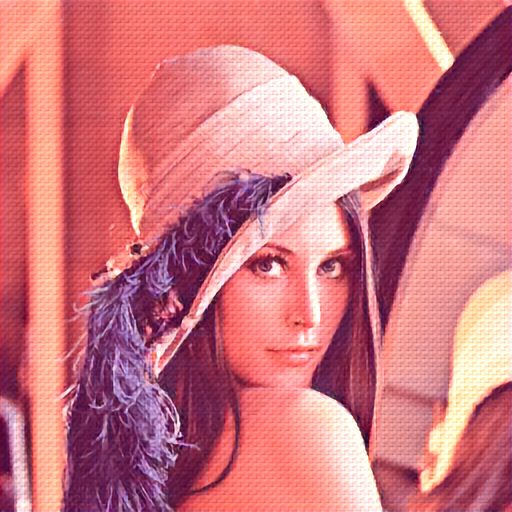}} \hspace{5mm}
\subfloat[][$\hat{\bm{\mathcal{X}}}$ with $i = 20$, PSNR = 32.56, SSIM = 0.933.]{\includegraphics[width=0.35\columnwidth]{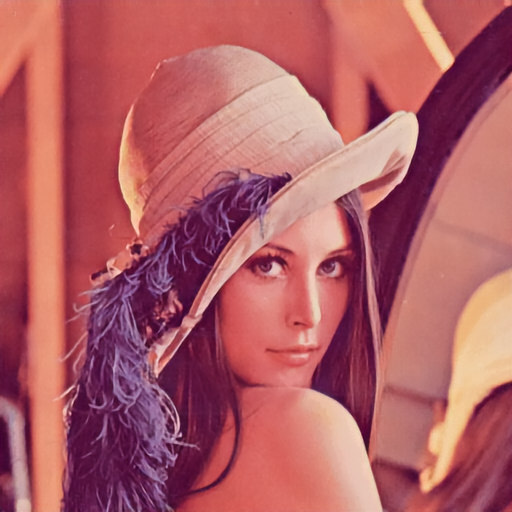}}  
\end{center}
\caption[caption]{Reconstruction at various recurrent latent optimization steps. The R\textsuperscript{2}CS-NET operates at 0.1 sampling rate and is trained with default recurrent steps = 20.} 
\label{fig: progressively optimization}
\end{figure}

We first validate the recurrent latent optimization design with ablation study. Sampling and reconstructing the \textit{Lena} image in Set14, Fig. \ref{fig: latent steps} illustrates the samplings' latent representation at various recurrent steps. The latent features are progressively optimized to accommodate the downstream reconstruction. Evolvement and convergence can be easily observed along the snapshots. Figure \ref{fig: progressively optimization} demonstrates the reconstructed image at each optimization step, evaluated under PSNR and structural similarity index measure (SSIM) metrics. The quality evolvement aligns with the progressive optimization principle and successfully verifies the architecture design.

\subsection{Sampling Efficiency and Reconstruction Quality}

\begin{table*}
\caption{Model Performance on LIVE1, Set5, Set14, BSD300, BSD500 Datasets\textsuperscript{*}: Using Averaged PSNR (dB) and SSIM Metrics (1st \& 2nd Column under each Dataset), and Time Cost (s) to Sample and Reconstruct All Images in LIVE1.\textsuperscript{+}}
\label{table: model performance}
 \setlength{\tabcolsep}{3 pt}
\centering
\begin{tabular}{|c|c|cc|cc|cc|cc|cc|c|c|}
\hline
\multirow{2}{*}{\makecell{Sampling\\ Rate}}  &  Model&  \multicolumn{10}{c|}{Dataset} & \multirow{2}{*}{\makecell{Time Cost \\ on LIVE1}} & \multirow{2}{*}{\makecell{Trainable \\ Parameters}} \\
\cline{3-12}
 & & \multicolumn{2}{c|}{LIVE1} &  \multicolumn{2}{c|}{Set5} & \multicolumn{2}{c|}{Set14} & \multicolumn{2}{c|}{BSD300} & \multicolumn{2}{c|}{BSD500} & &  \\
 \hline
 \multirow{5}{*}{0.02} & Random-RCS & 22.48 & 0.748 & 23.95 & 0.813 & 22.35 & 0.743  & 23.20 & 0.740  & 23.02 & 0.756 & 8.92 & 605,203 \\
 & RCS & 24.40 & 0.794 & 26.85 & 0.874 & 24.65 & 0.802  & 25.13 & 0.787  & 25.02 & 0.803 & 7.56  & 605,375 \\
 & RCS-GD-20 & 24.46 & \textcolor{blue}{\textbf{0.796}} & 26.98 & 0.876 & 24.74 & 0.804  & 25.21 & 0.789  & 25.09 & 0.805 & 206.82 & 605,375 \\
 & RCS-GD-100 & \textcolor{blue}{\textbf{24.52}} & \textcolor{red}{\textbf{0.798}} & \textcolor{red}{\textbf{27.10}} & \textcolor{red}{\textbf{0.879}} & \textcolor{red}{\textbf{24.82}} & \textcolor{red}{\textbf{0.806}}  & \textcolor{red}{\textbf{25.27}} & \textcolor{red}{\textbf{0.792}}  & \textcolor{red}{\textbf{25.16}} & \textcolor{red}{\textbf{0.808}}  & \textbf{635.66} & 605,375 \\
 & R\textsuperscript{2}CS-LSTM & 24.43 & 0.794 & 26.82 & 0.873 & 24.63 & 0.801  & 25.14 & 0.786  & 25.02 & 0.803 & 13.45 & 806,700 \\
  & R\textsuperscript{2}CS-NET & \textcolor{red}{\textbf{24.54}} & 0.794 & \textcolor{blue}{\textbf{26.99}} & \textcolor{blue}{\textbf{0.878}}  & \textcolor{blue}{\textbf{24.78}} & \textcolor{blue}{\textbf{0.805}}  & \textcolor{blue}{\textbf{25.25}} & \textcolor{blue}{\textbf{0.790}}  & \textcolor{blue}{\textbf{25.13}} & \textcolor{blue}{\textbf{0.806}} & \textbf{13.50}  & 1,028,581 \\
 \hline
 \multirow{5}{*}{0.05} & Random-RCS & 25.34 & 0.833 & 27.52 & 0.891 & 25.29 & 0.829  & 25.74 & 0.821  & 25.75 & 0.838 & 8.16 & 605,965 \\
 & RCS & 27.11 & 0.882 & 29.91 & 0.927 & 27.22 & 0.878  & 27.73 & 0.875  & 27.74 & 0.886 & 7.39 & 607,895 \\
 & RCS-GD-20 & 27.21 & 0.885 & 30.09 & 0.929 & 27.35 & 0.881  & 27.82 & 0.877  & 27.84 & 0.890 & 202.15 & 607,895 \\
 & RCS-GD-100 & \textcolor{blue}{\textbf{27.31}} & \textcolor{blue}{\textbf{0.887}} & \textcolor{blue}{\textbf{30.29}} & \textcolor{blue}{\textbf{0.932}} & \textcolor{blue}{\textbf{27.48}} & \textcolor{blue}{\textbf{0.884}}  & \textcolor{blue}{\textbf{27.92}} & \textcolor{red}{\textbf{0.881}}  & \textcolor{blue}{\textbf{27.95}} & \textcolor{red}{\textbf{0.893}} & \textbf{630.75} & 607,895 \\
 & R\textsuperscript{2}CS-LSTM & 27.25 & 0.885 & 30.05 & 0.929 & 27.39 & 0.882  & 27.75 & 0.876  & 27.80 & 0.890 & 13.23 & 811,668 \\
  & R\textsuperscript{2}CS-NET & \textcolor{red}{\textbf{27.50}} & \textcolor{red}{\textbf{0.889}} & \textcolor{red}{\textbf{30.33}} & \textcolor{red}{\textbf{0.933}}  & \textcolor{red}{\textbf{27.59}} & \textcolor{red}{\textbf{0.885}}  & \textcolor{red}{\textbf{27.97}} & \textcolor{blue}{\textbf{0.880}}  & \textcolor{red}{\textbf{28.03}} & \textcolor{red}{\textbf{0.893}} & \textbf{14.24}  & 1,033,099 \\
 \hline
 \multirow{5}{*}{0.10} & Random-RCS & 27.43 & 0.892 & 29.74 & 0.924 & 27.16 & 0.880  & 27.83 & 0.884  & 27.92 & 0.896 & 8.87 & 607,108 \\
 & RCS & 29.69 & 0.932 & 32.05 & 0.948 & 29.53 & 0.922  & 30.20 & 0.928  & 30.32 & 0.936 & 7.28 & 610,775 \\
 & RCS-GD-20 & 29.87 & 0.935 & 32.42 & 0.952 & 29.74 & 0.925  & 30.38 & 0.931  & 30.53 & 0.939 & 203.36 & 610,775 \\
 & RCS-GD-100 & 30.08 & \textcolor{blue}{\textbf{0.938}} & \textcolor{red}{\textbf{32.81}} & \textcolor{blue}{\textbf{0.956}} & \textcolor{blue}{\textbf{29.98}} & \textcolor{blue}{\textbf{0.927}}  & \textcolor{blue}{\textbf{30.59}} & \textcolor{blue}{\textbf{0.934}}  & \textcolor{blue}{\textbf{30.76}} & \textcolor{blue}{\textbf{0.942}} & \textbf{637.59}  & 610,775 \\
 & R\textsuperscript{2}CS-LSTM & \textcolor{blue}{\textbf{30.09}} & \textcolor{blue}{\textbf{0.938}} & 32.54 & 0.954 & 29.91 & \textcolor{blue}{\textbf{0.927}}  & 30.54 & 0.933  & 30.70 & 0.942 & 13.14 & 821,145 \\
  & R\textsuperscript{2}CS-NET & \textcolor{red}{\textbf{30.31}} & \textcolor{red}{\textbf{0.940}} & \textcolor{blue}{\textbf{32.79}} & \textcolor{red}{\textbf{0.957}}  & \textcolor{red}{\textbf{30.09}} & \textcolor{red}{\textbf{0.929}}  & \textcolor{red}{\textbf{30.72}} & \textcolor{red}{\textbf{0.935}}  & \textcolor{red}{\textbf{30.89}} & \textcolor{red}{\textbf{0.943}} & \textbf{13.39} & 1,043,116 \\
 \hline
 \multirow{5}{*}{0.20} & Random-RCS & 30.66 & 0.946 & 32.45 & 0.951 & 29.69 & 0.922  & 31.25 & 0.946  & 31.32 & 0.950 &  8.81 & 609,648 \\
 & RCS & 33.97 & 0.973 & 35.35 & 0.970 & 32.82 & 0.955  & 34.74 & 0.974  & 34.77 & 0.976 & 7.32 & 617,175 \\
 & RCS-GD-20 & 34.33 & 0.975 & 35.82 & 0.973 & 33.15 & 0.958  & 35.14 & 0.976  & 35.19 & 0.978 & 207.00 & 617,175 \\
 & RCS-GD-100 & \textcolor{blue}{\textbf{34.75}} & \textcolor{red}{\textbf{0.978}} & \textcolor{red}{\textbf{36.43}} & \textcolor{red}{\textbf{0.976}} & \textcolor{blue}{\textbf{33.54}} & 0.960  & \textcolor{red}{\textbf{35.62}} & \textcolor{red}{\textbf{0.979}}  & \textcolor{red}{\textbf{35.69}} & \textcolor{red}{\textbf{0.980}} & \textbf{639.78} & 617,175 \\
 & R\textsuperscript{2}CS-LSTM & 34.67 & \textcolor{red}{\textbf{0.978}} & 35.90 & \textcolor{blue}{\textbf{0.974}} & 33.47 &  \textcolor{red}{\textbf{0.961}}  & 35.34 & \textcolor{blue}{\textbf{0.978}}  & 35.42 & \textcolor{red}{\textbf{0.980}} & 13.25 & 840,405 \\
  & R\textsuperscript{2}CS-NET &  \textcolor{red}{\textbf{34.79}} & \textcolor{red}{\textbf{0.978}} & \textcolor{blue}{\textbf{35.97}} & \textcolor{blue}{\textbf{0.974}}  &  \textcolor{red}{\textbf{33.56}} & \textcolor{red}{\textbf{0.961}}  & \textcolor{blue}{\textbf{35.54}} & \textcolor{blue}{\textbf{0.978}}  & \textcolor{blue}{\textbf{35.59}} & \textcolor{red}{\textbf{0.980}} & \textbf{13.62} & 1,079,296 \\
 \hline
  \multirow{5}{*}{0.30} & Random-RCS & 33.46 & 0.971 & 34.92 & 0.966 & 31.92 & 0.944  & 34.36 & 0.972  & 34.37 & 0.974 & 8.83 & 612,061 \\
  & RCS & 37.07 & 0.984 & 37.45 & 0.977 & 35.05 & 0.967  & 37.83 & 0.986  & 37.79 & 0.986 & 7.41 & 623,255 \\
  & RCS-GD-20 & 37.78 & 0.987 & 38.22 & 0.981 & 35.65 & 0.970  & 38.67 & 0.989  & 38.60 & 0.989 & 217.13 & 623,255 \\
 & RCS-GD-100 & \textcolor{blue}{\textbf{38.68}} & \textcolor{blue}{\textbf{0.990}} & \textcolor{red}{\textbf{39.33}} & \textcolor{red}{\textbf{0.984}} & \textcolor{blue}{\textbf{36.45}} & \textcolor{blue}{\textbf{0.974}}  & \textcolor{blue}{\textbf{39.74}} & \textcolor{blue}{\textbf{0.991}}  & \textcolor{blue}{\textbf{39.65}} & \textcolor{blue}{\textbf{0.991}} & \textbf{645.36} & 623,255 \\
 & R\textsuperscript{2}CS-LSTM & 37.98 & 0.989 & 38.36 & 0.982 & 35.97 & 0.973  & 39.10 & 0.990  & 38.96 & 0.990 & 13.14 & 858,702 \\
 & R\textsuperscript{2}CS-NET & \textcolor{red}{\textbf{38.90}} & \textcolor{red}{\textbf{0.991}} & \textcolor{blue}{\textbf{39.07}} & \textcolor{red}{\textbf{0.984}}  & \textcolor{red}{\textbf{38.90}} & \textcolor{red}{\textbf{0.991}}  & \textcolor{red}{\textbf{39.87}} & \textcolor{red}{\textbf{0.992}}  & \textcolor{red}{\textbf{39.73}} & \textcolor{red}{\textbf{0.992}} & \textbf{13.54} & 1,131,451 \\
 \hline
 \multicolumn{14}{l}{*LIVE1, Set5, Set14, and test sets of BSD300, BSD500 contain 29, 5, 14, 100, and 200 color images, respectively.} \\
 \multicolumn{14}{l}{+The best and second best results are highlighted in \textcolor{red}{\textbf{red}} and \textcolor{blue}{\textbf{blue}}, respectively. Their time costs are highlighted in \textbf{bold}. } \\
  
\end{tabular}
\end{table*}

The efficiency of learnable measurement matrix, the effectiveness of the residual CNN reconstruction, and the contribution of recurrent latent optimization are further validated through lateral comparison with control group counterparts. Each of the following architectures is investigated on NVIDIA GeForce RTX 2080 Super GPU with the same runtime setting.
\begin{itemize}
  \item Random-RCS: samples with fixed, classic random measurement matrix; reconstructs via the residual CNN without online optimization.
  \item RCS: samples with the learnable measurement matrix, with remaining parts the same as the Random-RCS.
  \item RCS-GD: employs the gradient descent adaptive online optimization upon the RCS, following Eq. (\ref{eq: latent space reconstruction}). Iteration steps of 20 and 100 are investigated in the experiment, where convergence is empirically reached at 100 steps.
  \item R\textsuperscript{2}CS-LSTM: employs the classic LSTM architecture for recurrent latent optimization upon the RCS. The LSTM unit passes measurement residuals in the same manner as the proposed recurrent-residual stage.
  \item R\textsuperscript{2}CS-NET: the proposed framework.
\end{itemize}
Five groups of model instances are trained at 0.02, 0.05, 0.1, 0.2 and 0.3 sampling rate, respectively. Their statistical performance is evaluated on five widely used benchmark datasets: LIVE1 \cite{1709988}, Set5  \cite{Timofte_2018_CVPR_Workshops}, Set14 \cite{Timofte_2018_CVPR_Workshops}, and the test sets of BSD300 \cite{MartinFTM01}, BSD500 \cite{amfm_pami2011}.

Table \ref{table: model performance} summarizes the performance of each experimental group. The Random-RCS serves as the base case. Its samplings are acquired through random measurement matrix with no online optimization. The successful reconstruction, meanwhile, demonstrates the fundamental validity of the residual convolutional reconstruction stage. Comparison between the Random-RCS and the RCS reveals the sampling efficiency delivered by the learnable measurement matrix. The CS methods with online optimization, namely the RCS-GD, R\textsuperscript{2}CS-LSTM, and R\textsuperscript{2}CS-NET, demonstrate better performance in general. The consistently superior reconstruction of the RCS-GD over the RCS indicates the significance of online optimization. Despite the RCS-GD's leading CS fidelity, its classic gradient descent online optimizer is extremely slow. The R\textsuperscript{2}CS-NET employs the proposed recurrent latent optimization stage to achieve comparable reconstruction quality with considerably higher efficiency. Its time efficiency and slightly better fidelity over the RCS-GD-100 attest the effectiveness. As a control group, the R\textsuperscript{2}CS-LSTM utilizes the classic convolutional LSTM unit for adaptive optimization. The proposed recurrent architecture in the R\textsuperscript{2}CS-NET outperforms the LSTM in the same scenario. Though the R\textsuperscript{2}CS-NET contains more parameters, the two methods operate at comparable speed.

\begin{figure*}[!htb]
\begin{center}
\captionsetup[subfigure]{justification=centering,labelformat=empty,position=top}
\subfloat[][Ground Truth \\ Noise $\sigma$]{\includegraphics[width=0.11\textwidth]{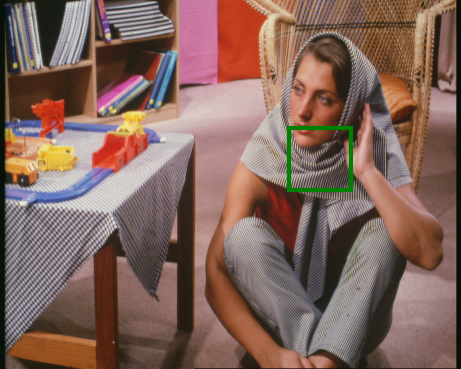}}  
\subfloat[][Random-RCS \\ Clean]{\includegraphics[width=0.111\textwidth]{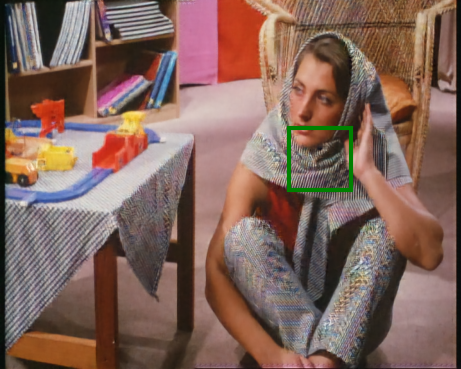}}  
\subfloat[][RCS \\ Clean]{\includegraphics[width=0.111\textwidth]{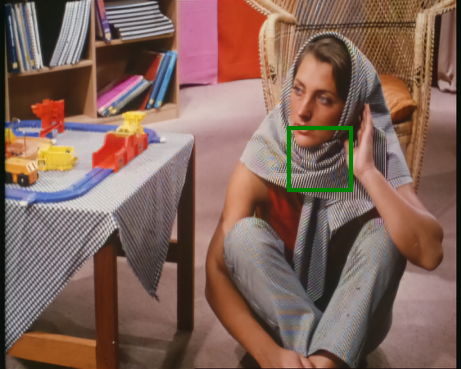}}  
\subfloat[][RCS-GD-100 \\ Clean]{\includegraphics[width=0.111\textwidth]{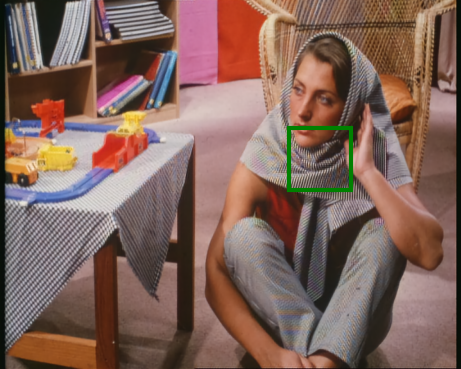}}  
\subfloat[][R\textsuperscript{2}CS-NET \\ Clean]{\includegraphics[width=0.111\textwidth]{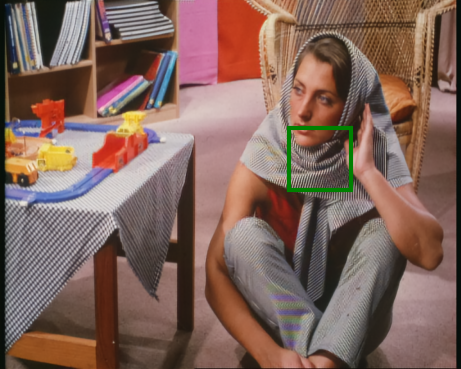}}  
\subfloat[][Random-RCS \\ $\sigma = 0.3$]{\includegraphics[width=0.111\textwidth]{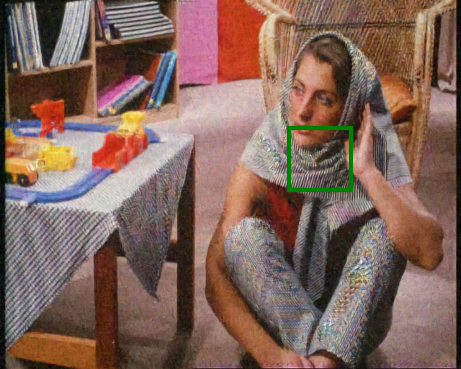}}  
\subfloat[][RCS \\ $\sigma = 0.3$]{\includegraphics[width=0.111\textwidth]{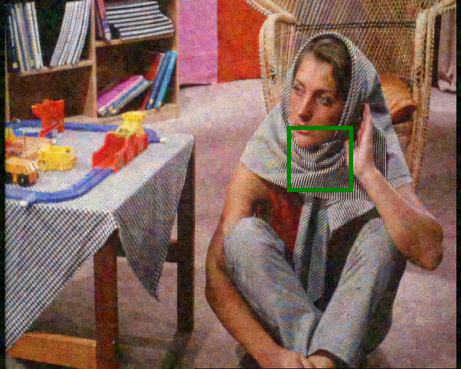}}  
\subfloat[][RCS-GD-100 \\ $\sigma = 0.3$]{\includegraphics[width=0.111\textwidth]{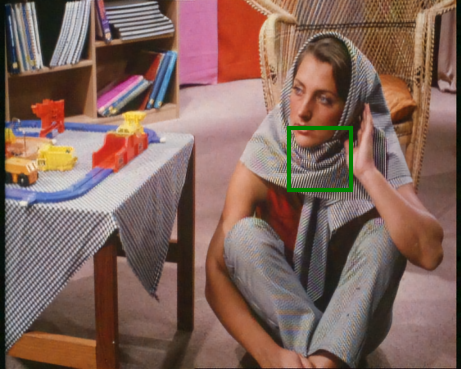}}  
\subfloat[][R\textsuperscript{2}CS-NET \\ $\sigma = 0.3$]{\includegraphics[width=0.111\textwidth]{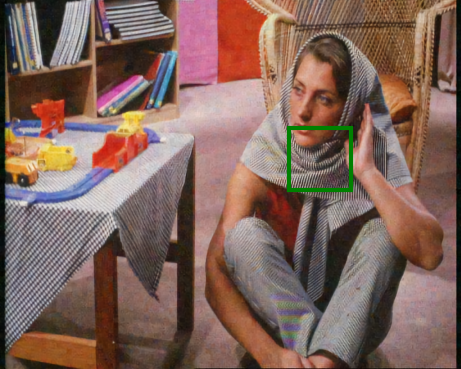}}  
\\[-2ex]
\noindent 
\captionsetup[subfigure]{justification=centering,labelformat=empty,position=bottom}
\subfloat[][PNSR/SSIM]{\includegraphics[width=0.11\textwidth]{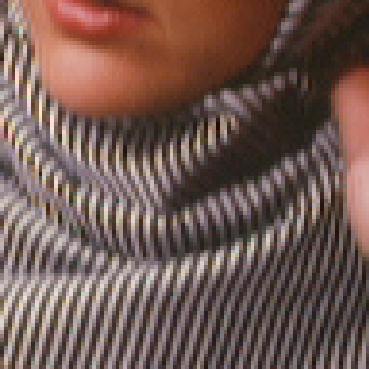}}  
\subfloat[][25.85/0.876]{\includegraphics[width=0.11\textwidth]{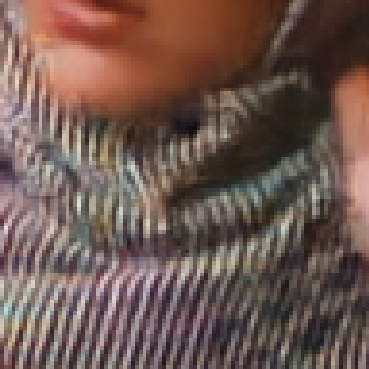}}   
\subfloat[][27.23/0.895]{\includegraphics[width=0.11\textwidth]{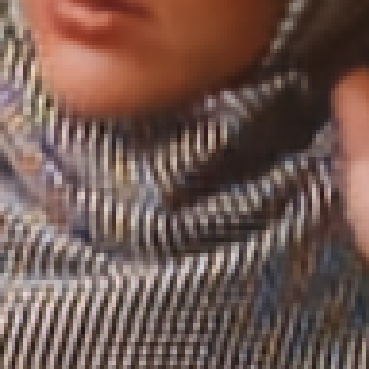}}   
\subfloat[][27.39/0.900]{\includegraphics[width=0.11\textwidth]{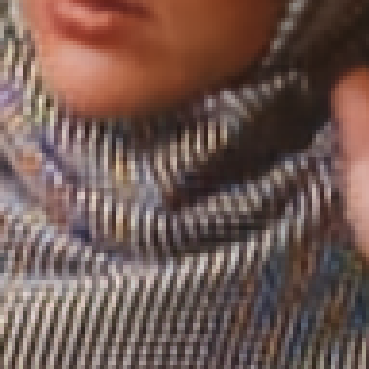}}   
\subfloat[][27.41/0.902]{\includegraphics[width=0.11\textwidth]{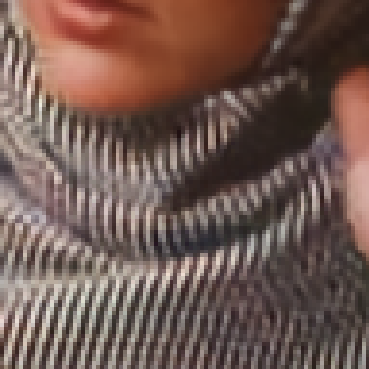}}  
\subfloat[][24.74/0.817]{\includegraphics[width=0.11\textwidth]{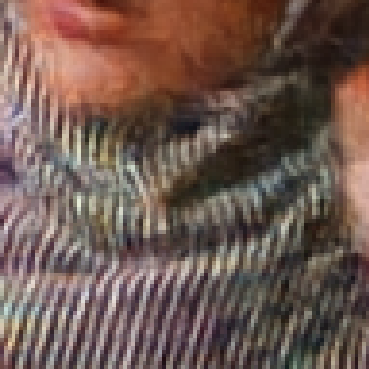}}   
\subfloat[][25.02/0.812]{\includegraphics[width=0.11\textwidth]{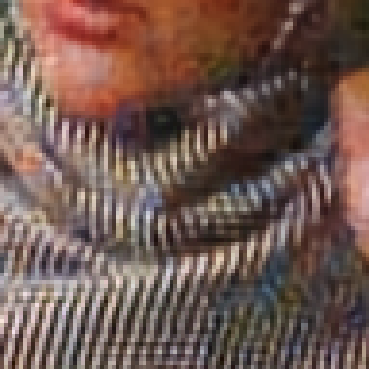}}   
\subfloat[][27.15/0.892]{\includegraphics[width=0.11\textwidth]{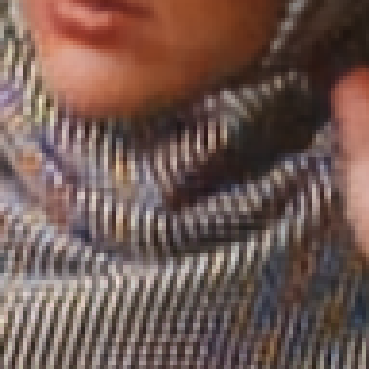}}   
\subfloat[][26.49/0.873]{\includegraphics[width=0.11\textwidth]{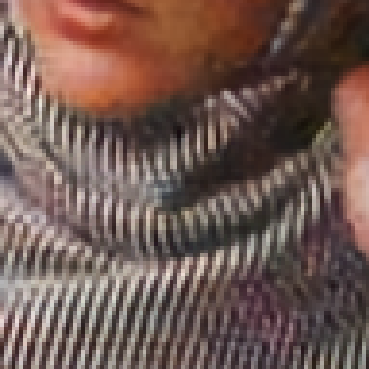}}   

\end{center}
\caption[caption]{Reconstructions on \textit{Barbara} image in Set14 at sampling rate of 0.1. Highlighted regions are enlarged in the 2nd row.} 
\label{fig: reconstruction examples}
\end{figure*}

Figure \ref{fig: reconstruction examples} shows the reconstructions of four methods on \textit{Barbara} in Set14, from both clean and noisy measurement environment. The R\textsuperscript{2}CS-NET achieves comparable fidelity as the RCS-GD-100, prevailing the methods absenting online optimization. The recurrent-residual structural constraint introduces moderate latency in adaptive optimization. However, decoupled from sampling, it has no direct impact on measurement latency or sensor hardware complexity.

\subsection{Robustness under Measurement Noise}

The recurrent latent optimization conceptually inherits the robustness of online optimization. The potential is further realized through measurement perturbation during training.  This experiment investigates the R\textsuperscript{2}CS-NET's robustness under various measurement noise levels. 

\begin{figure}[ht]
\begin{center}
\subfloat[][Reconstruction quality comparison across progressive latent optimization architectures. \label{fig: noise vs rnns}]{\includegraphics[width=0.7\columnwidth]{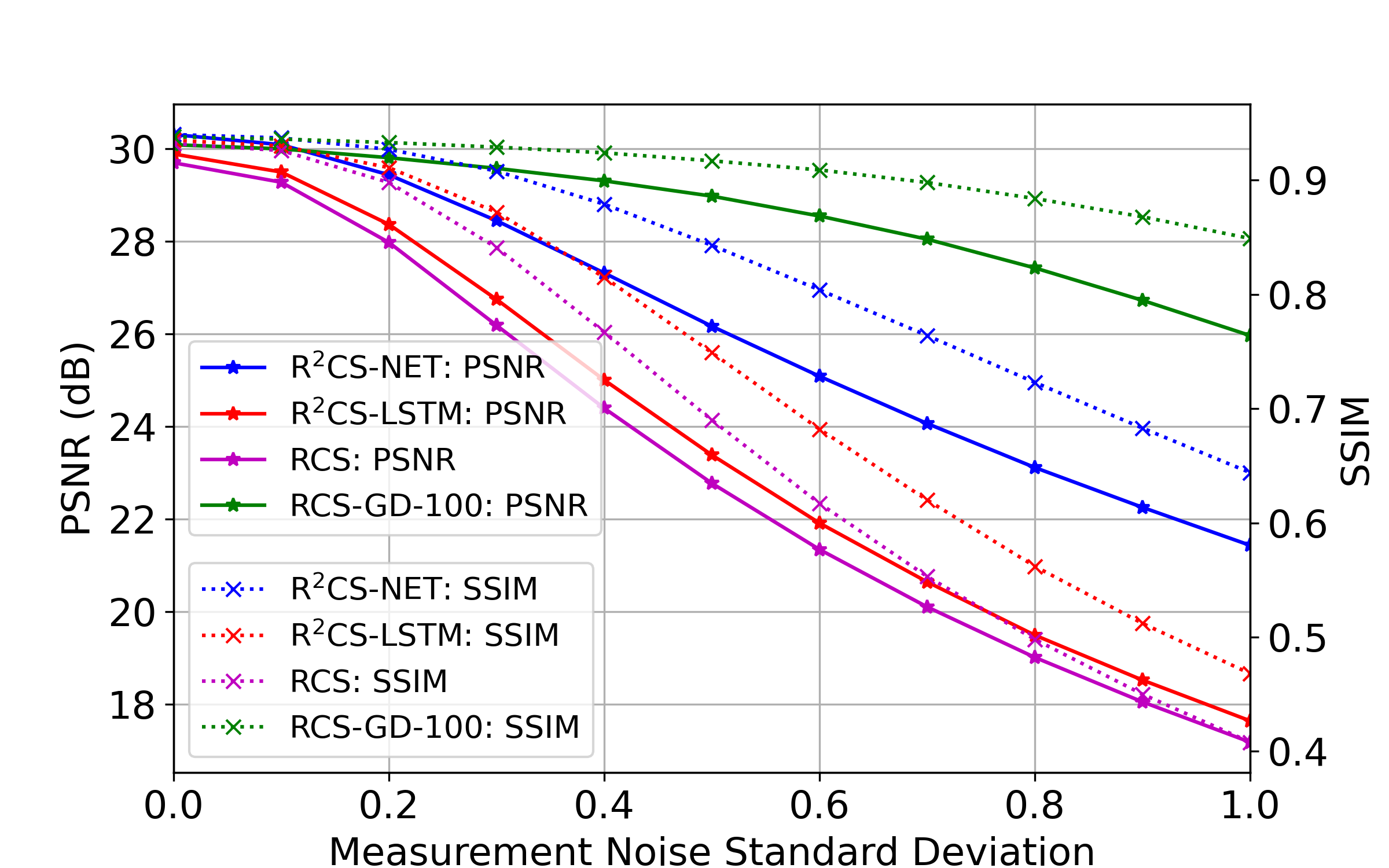}} \hfill 
\subfloat[][Measurement noise robustness comparison with reported benchmarks. \label{fig: noise vs benchmarks}]{\includegraphics[width=0.7\columnwidth]{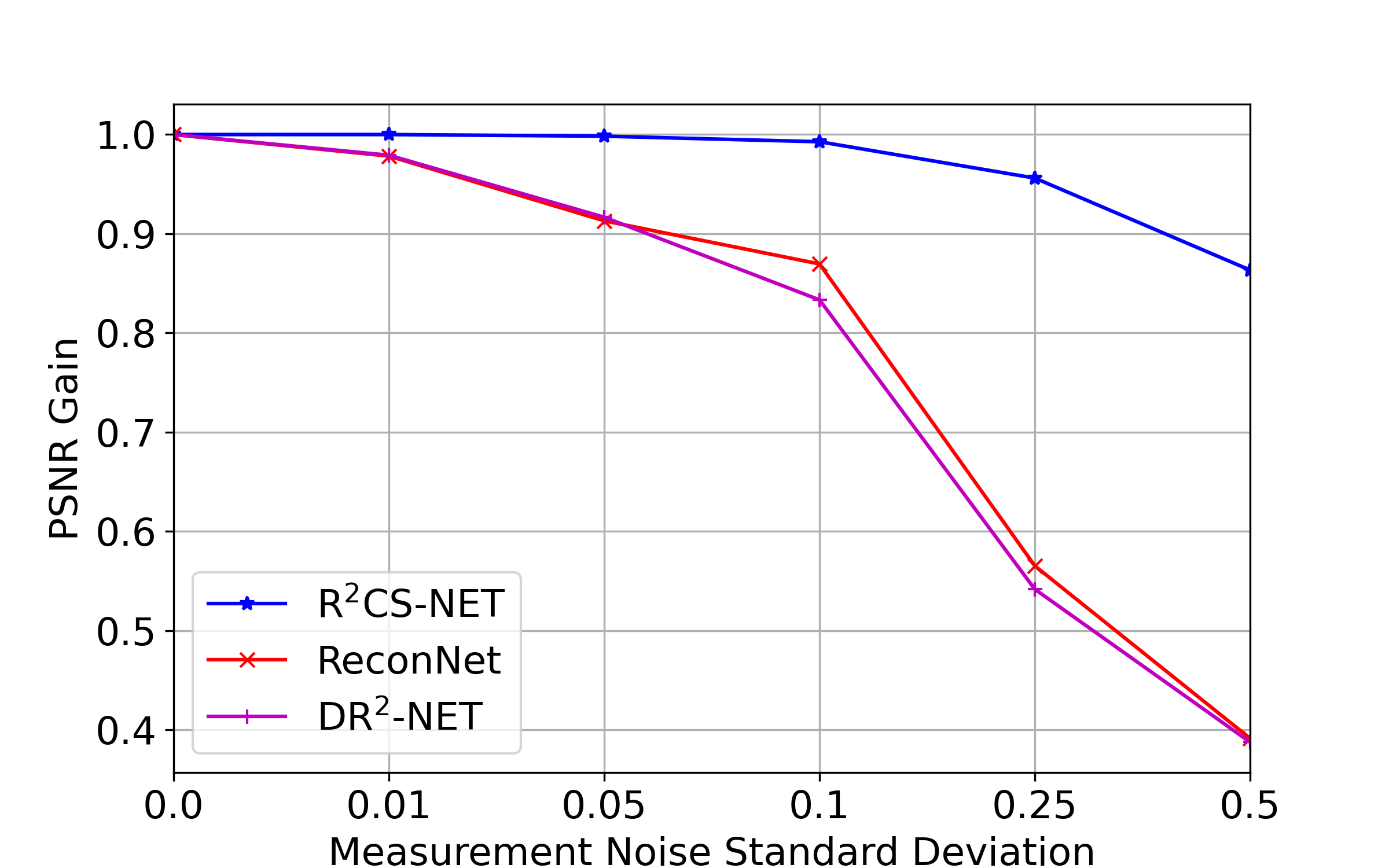}}
\end{center}
\caption[caption]{Comparison between R\textsuperscript{2}CS-NET and benchmarks under various measurement noise levels at 0.1 sampling rate.} 
\label{fig: noise robustness}
\end{figure}

To demonstrate the noise robustness contributed by the recurrent-residual structural constraint, we again compare the R\textsuperscript{2}CS-NET with the RCS, the RCS-GD-100, and the the R\textsuperscript{2}CS-LSTM. Figure \ref{fig: noise vs rnns} compares the four architectures under measurement noise level from $\sigma = 0$ to $\sigma = 1.0$. The RCS-GD-100 unsurprisingly shows highest robustness. Its iterative online optimization amends the latent representation at the trade-off of computational cost. Meanwhile, the R\textsuperscript{2}CS-NET gains stronger noise robustness over the RCS and the R\textsuperscript{2}CS-LSTM. The margin expands along the rising measurement noise level. The proposed recurrent unit outperforms the LSTM regarding noise robustness as well.

Figure \ref{fig: noise vs benchmarks} provides the R\textsuperscript{2}CS-NET's measurement noise sensitivity versus the benchmarks reported by the DR\textsuperscript{2}-NET \cite{YAO2019483} and the ReconNet  \cite{8379450}. Based on the available benchmark data, the models are compared at $\sigma= 0, 0.01, 0.05, 0.1, 0.25, 0.5$, with 0.1 sampling rate. Under all noise levels, the R\textsuperscript{2}CS-NET demonstrates considerably higher robustness. The margin expands in higher noise region. For example, the DR\textsuperscript{2}-NET and the ReconNet's reconstruction quality drops by 61\% at $\sigma=0.5$ noise level. On the contrary, the R\textsuperscript{2}CS-NET is only impaired by 13\%.

Comparison between the R\textsuperscript{2}CS-NET and the RCS reveals the recurrent latent optimization's direct contribution against noise. Fig. \ref{fig: noise robustness} visualizes the two methods' reconstructions on \textit{Monarch}. Enhancements are observed at all noise levels. 

\begin{figure*}[!htb]
\begin{center}
\captionsetup[subfigure]{justification=centering,labelformat=empty,position=top}
\subfloat[][Ground Truth \\ Noise $\sigma$]{\includegraphics[width=0.11\textwidth]{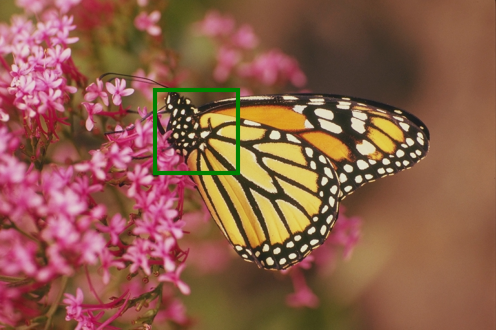}}  
\subfloat[][RCS \\ Clean]{\includegraphics[width=0.111\textwidth]{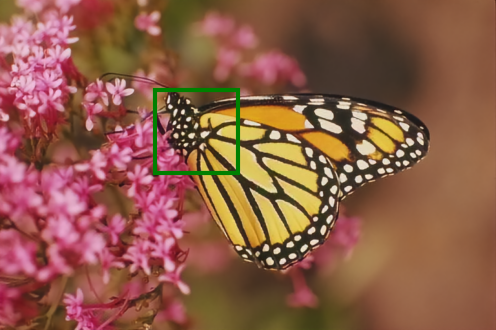}}  
\subfloat[][RCS \\ $\sigma = 0.1$]{\includegraphics[width=0.111\textwidth]{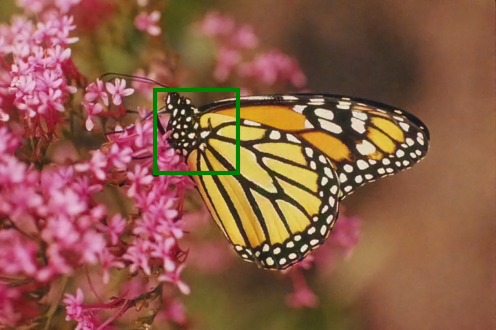}}  
\subfloat[][RCS \\ $\sigma = 0.3$]{\includegraphics[width=0.111\textwidth]{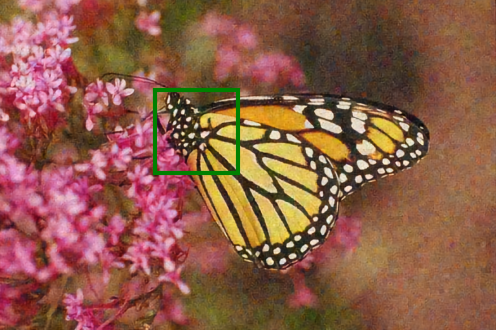}}  
\subfloat[][RCS \\ $\sigma = 0.5$]{\includegraphics[width=0.111\textwidth]{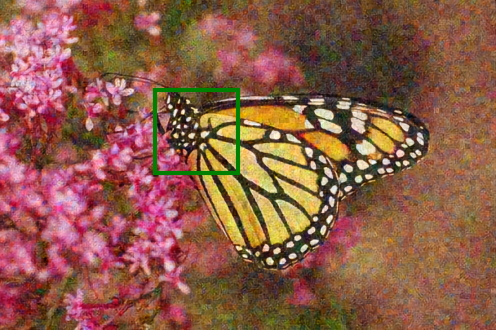}}  
\subfloat[][R\textsuperscript{2}CS-NET \\ Clean]{\includegraphics[width=0.111\textwidth]{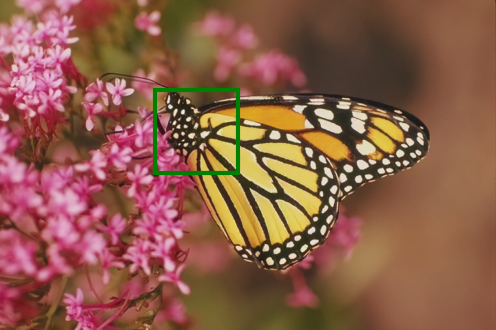}}  
\subfloat[][R\textsuperscript{2}CS-NET \\ $\sigma = 0.1$]{\includegraphics[width=0.111\textwidth]{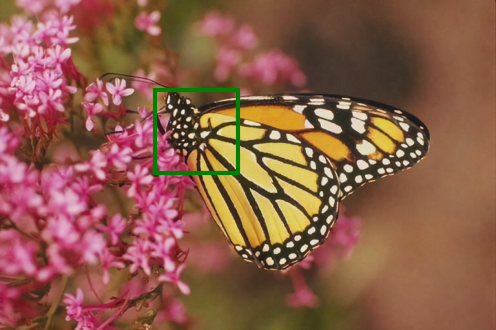}}  
\subfloat[][R\textsuperscript{2}CS-NET \\ $\sigma = 0.3$]{\includegraphics[width=0.111\textwidth]{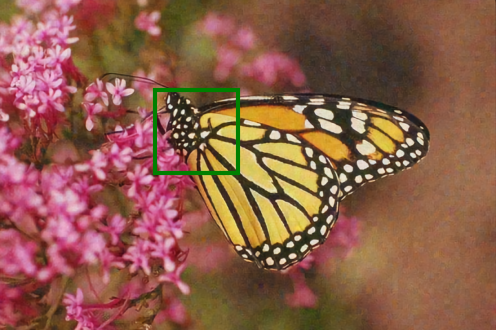}}  
\subfloat[][R\textsuperscript{2}CS-NET \\ $\sigma = 0.5$]{\includegraphics[width=0.111\textwidth]{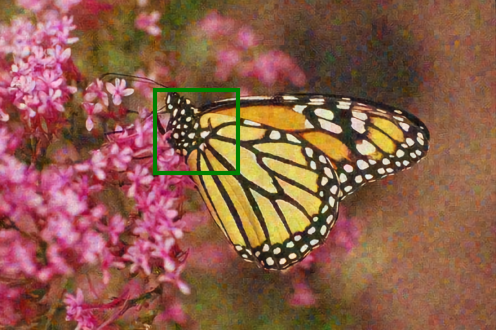}}  
\\[-2ex]
\noindent 
\captionsetup[subfigure]{justification=centering,labelformat=empty,position=bottom}
\subfloat[][PNSR/SSIM]{\includegraphics[width=0.11\textwidth]{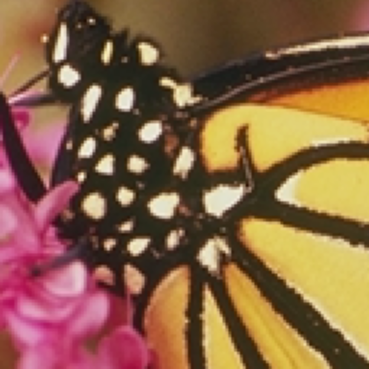}}  
\subfloat[][32.68/0.972]{\includegraphics[width=0.11\textwidth]{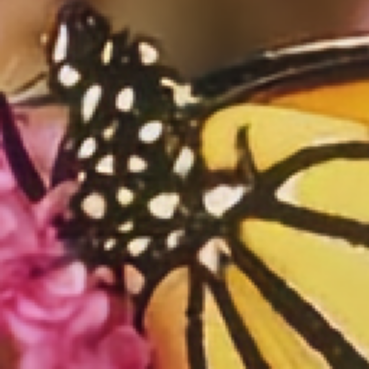}}   
\subfloat[][32.05/0.966]{\includegraphics[width=0.11\textwidth]{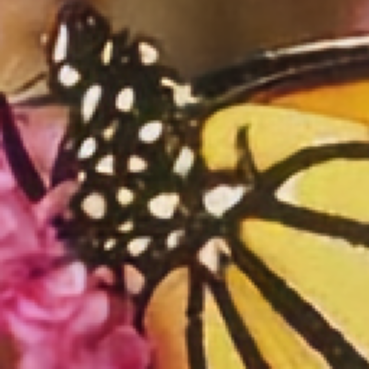}}   
\subfloat[][27.63/0.867]{\includegraphics[width=0.11\textwidth]{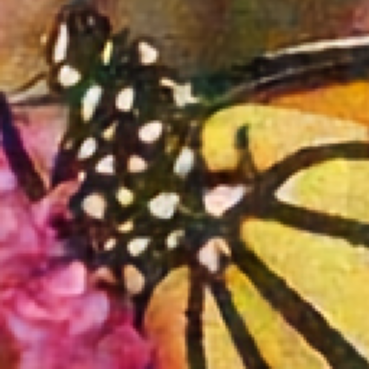}}   
\subfloat[][23.44/0.690]{\includegraphics[width=0.11\textwidth]{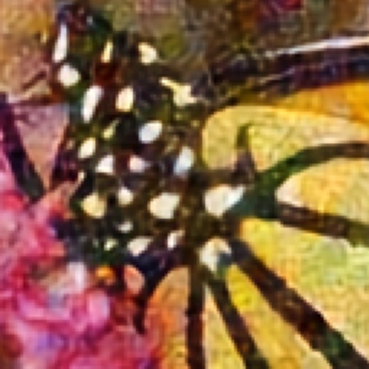}}  
\subfloat[][33.96/0.977]{\includegraphics[width=0.11\textwidth]{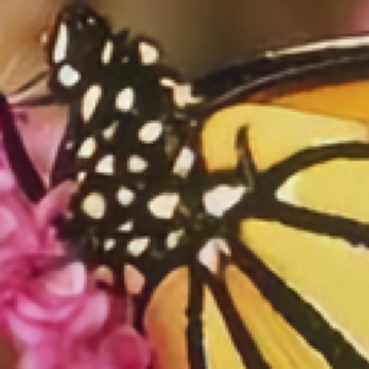}}   
\subfloat[][33.59/0.974]{\includegraphics[width=0.11\textwidth]{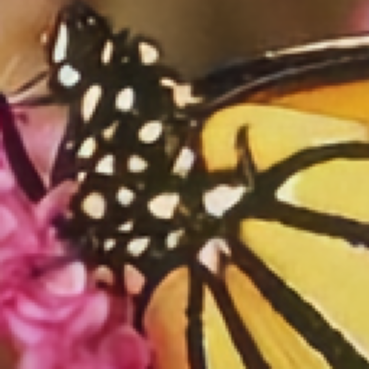}}   
\subfloat[][30.95/0.942]{\includegraphics[width=0.11\textwidth]{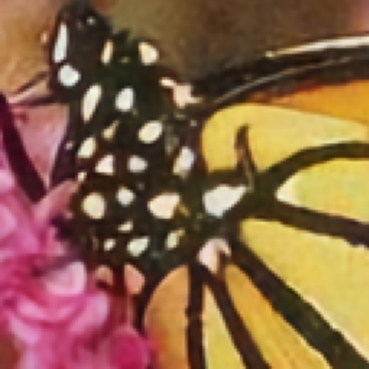}}   
\subfloat[][27.65/0.866]{\includegraphics[width=0.11\textwidth]{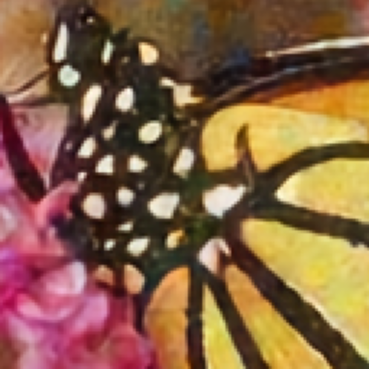}}   

\end{center}
\caption[caption]{R\textsuperscript{2}CS-NET and RCS reconstruction on \textit{Monarch} image in Set14 at sampling rate of 0.1, under various measurement noise levels. Highlighted regions are enlarged in the 2nd row.} 
\label{fig: noise robustness}
\end{figure*}

\subsection{Comparing with Deep CS Benchmarks}
Verified the R\textsuperscript{2}CS-NET's validity via stage-by-stage lateral comparisons, this section further investigates the overall framework's performance amidst recent deep CS benchmarks. The R\textsuperscript{2}CS-NET enlists its efficacy through learnable sampling and efficient adaptive reconstruction. The correlations across and within image channels are carefully addressed at both measurement and reconstruction stages. Additionally assisted with recurrent latent optimization, the R\textsuperscript{2}CS-NET's adaptive sensing nature makes it a natural candidate for color image CS. Table \ref{table: compare BSD500} first compares it with existing benchmarks on BSD500. Noticing most existing CS benchmarks are reported on grayscale images, a set of R\textsuperscript{2}CS-NET instances are further trained with single channel images. Table \ref{table: compare set 11}  inspects the candidates on Set11 \cite{9159912}, containing 11 grayscale images. Among the benchmarks, TVAL3 \cite{RePEc:spr:coopap:v:56:y:2013:i:3:p:507-530}, DAMP \cite{7457256} are classic model-based CS methods.  LDIT \cite{10.5555/3294771.3294940}, LDAMP \cite{10.5555/3294771.3294940} are based on random measurement, while the remaining methods generally support learning measurement matrix through data. ReconNet \cite{8379450} and CSNet\textsuperscript{+} \cite{8765626} employ classic deep network architecture, while AMP-NET \cite{9298950}, LDIT, LDAMP, NN \cite{8878159}, DPDNN \cite{8481558}, ISTA-NET\textsuperscript{+} \cite{8578294} unfold the iterative optimization into feed-forward networks. CSNet\textsuperscript{+}, LDIT, LDAMP fully address color images' channel correlation. R\textsuperscript{2}CS-NET is the first method integrates adaptive online optimization with unrolled network.

\begin{table}
\caption{Model Comparison on the Color BSD500 Test Set, Using Averaged PSNR (dB) and SSIM Metrics (1st  \& 2nd Column under each Sampling Rate).}
\label{table: compare BSD500}
 \setlength{\tabcolsep}{3 pt}
\centering
\begin{tabular}{|c|c|cc|cc|cc|}
  \hline
  Dataset&  Model&  \multicolumn{6}{c|}{Sampling Rate}  \\
  \cline{3-8}
   &  & \multicolumn{2}{c|}{0.1} &  \multicolumn{2}{c|}{0.25} & \multicolumn{2}{c|}{0.3} \\
  \hline
\multirow{9}{*}{\makecell{BSD500 \\ Test Set}} &  ReconNet \cite{8379450} & 26.85 & 0.784  & 30.51 & 0.897  & 31.49 & 0.917  \\
 &  CSNet\textsuperscript{+} \cite{8765626} & 29.07 & 0.820  & 33.26 & 0.925  & 34.52 & 0.945 \\
 &  LDIT \cite{10.5555/3294771.3294940} & 24.94 & 0.704  & 29.10 & 0.852  & 30.23 & 0.880 \\
 &  LDAMP \cite{10.5555/3294771.3294940} & 26.61 & 0.692  & 28.35 & 0.830  & 29.89 & 0.872 \\
 &  DPDNN \cite{8481558} & 24.37 & 0.686  & 28.87 & 0.849  & 29.98 & 0.876 \\
 &  NN \cite{8878159} & 23.44 & 0.644  & 26.42 & 0.776  & 27.23 & 0.804 \\
 &  ISTA-NET\textsuperscript{+}  \cite{8578294} & 25.46 & 0.502  & 29.92 & 0.738  & 32.21 & 0.769 \\
 &  AMP-NET \cite{9298950} & 28.11 & 0.810  & 32.55 & 0.923  & 33.82 & 0.941 \\
 
 &  R\textsuperscript{2}CS-NET & \textbf{30.89} & \textbf{0.943} & \textbf{37.56} & \textbf{0.987}  & \textbf{39.73} & \textbf{0.992} \\

\hline
\end{tabular}
\end{table}

\begin{figure}[!htb]
\begin{center}
\captionsetup[subfigure]{justification=centering,labelformat=empty,position=top}
\subfloat[][Ground Truth]{\includegraphics[width=0.1\textwidth]{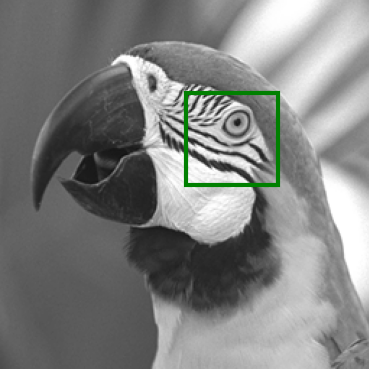}}  
\subfloat[][ReconNet]{\includegraphics[width=0.1\textwidth]{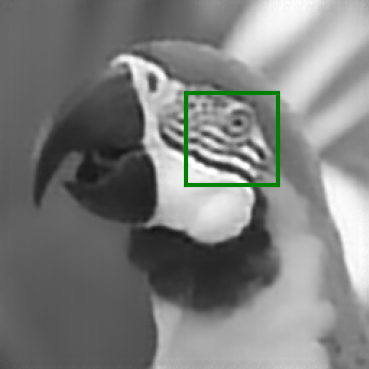}}  
\subfloat[][LDAMP]{\includegraphics[width=0.1\textwidth]{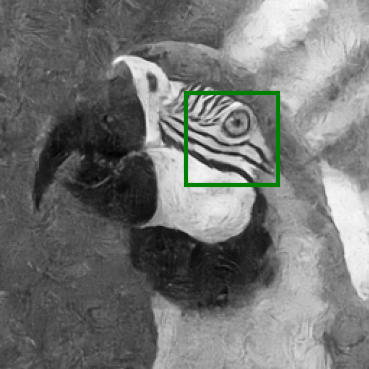}}  
\subfloat[][LDIT]{\includegraphics[width=0.1\textwidth]{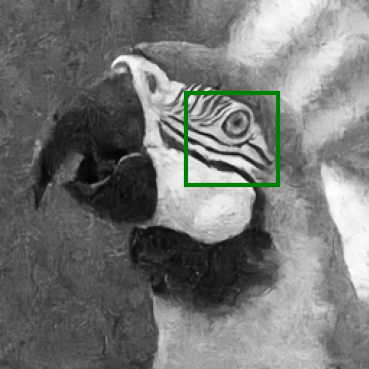}}  
\\[-2ex]
\noindent 
\captionsetup[subfigure]{justification=centering,labelformat=empty,position=bottom}
\subfloat[][PNSR/SSIM]{\includegraphics[width=0.1\textwidth]{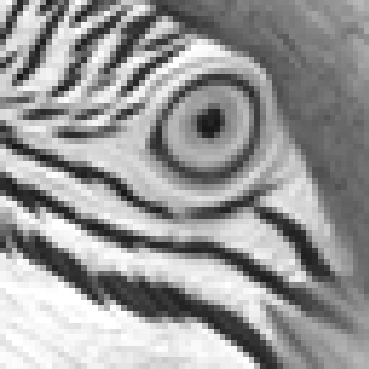}}  
\subfloat[][25.56/0.902]{\includegraphics[width=0.1\textwidth]{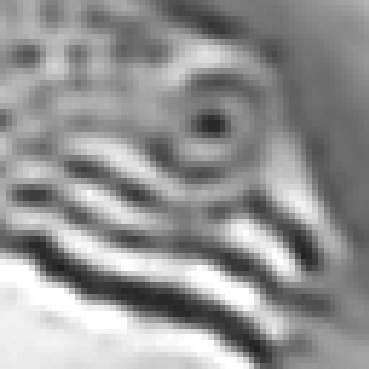}}  
\subfloat[][25.55/0.783]{\includegraphics[width=0.1\textwidth]{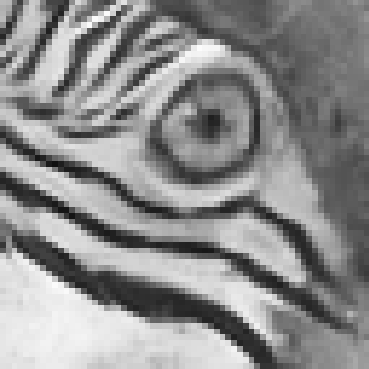}}  
\subfloat[][25.99/0.799]{\includegraphics[width=0.1\textwidth]{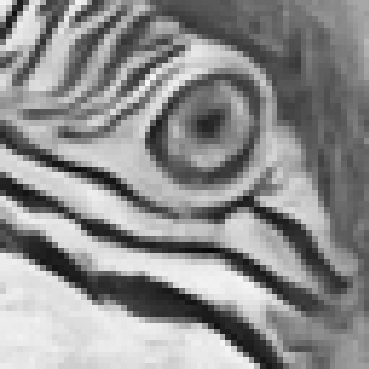}}  
\\[-1ex]
\noindent 
\captionsetup[subfigure]{justification=centering,labelformat=empty,position=top}
\subfloat[][CSNet\textsuperscript{+}]{\includegraphics[width=0.1\textwidth]{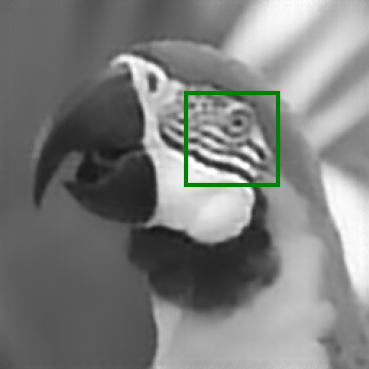}}  
\subfloat[][AMP-NET]{\includegraphics[width=0.1\textwidth]{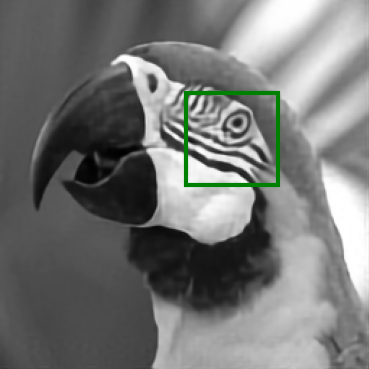}}  
\subfloat[][R\textsuperscript{2}CS-NET]{\includegraphics[width=0.1\textwidth]{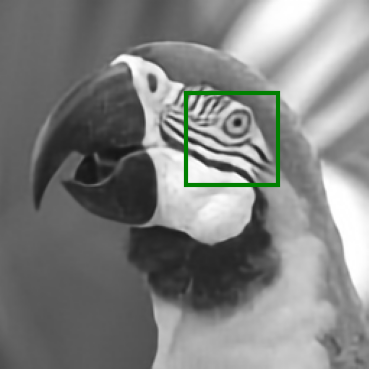}}  
\subfloat[][R\textsuperscript{2}CS-NET\textsubscript{k}]{\includegraphics[width=0.1\textwidth]{Fig_10_R2CS_NET_rected.png}}  
\\[-2ex]
\noindent 
\captionsetup[subfigure]{justification=centering,labelformat=empty,position=bottom}
\subfloat[][28.11/0.890]{\includegraphics[width=0.1\textwidth]{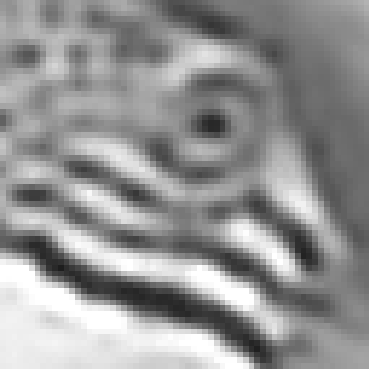}}  
\subfloat[][28.93/0.903]{\includegraphics[width=0.1\textwidth]{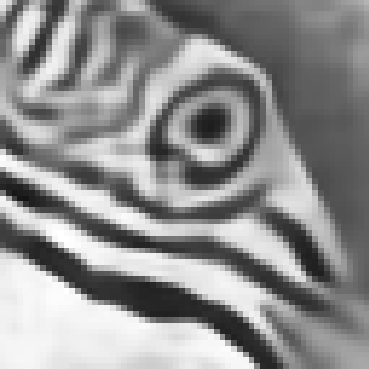}}  
\subfloat[][29.62/0.955]{\includegraphics[width=0.1\textwidth]{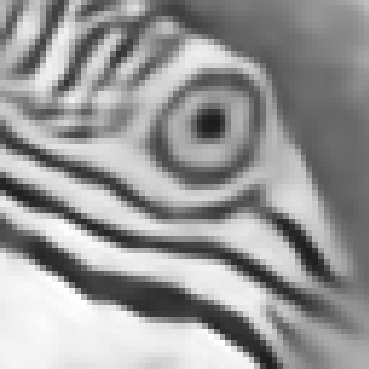}}  
\subfloat[][29.62/0.955]{\includegraphics[width=0.1\textwidth]{Fig_10_R2CS_NET_detail_0.png}}  
\end{center}
\caption[caption]{Reconstructions on \textit{Parrots} image in Set11 at 0.1 sampling rate. R\textsuperscript{2}CS-NET\textsubscript{k} acquires samplings in k-space; all others sample from pixel domain.} 
\label{fig: benchmark comparison}
\end{figure}

The adaptive online optimization as well as the R\textsuperscript{2}CS-NET's overall architecture successfully handles sampling correlation, contributing toward outstanding performance in color image CS. The margin over benchmark methods expands along with the sampling rate. Leading performance can also be observed on grayscale test. Figure \ref{fig: benchmark comparison} visualizes reconstructions on \textit{Parrots} image in Set11 at 0.1 sampling rate. All methods sample pixel domain images except the R\textsuperscript{2}CS-NET\textsubscript{k}. As a validation of the framework's sampling adaptivity, the R\textsuperscript{2}CS-NET\textsubscript{k} transplants the R\textsuperscript{2}CS-NET for k-space sampling by merely switching the measurement mask. It achieves the same outcome except for inevitable rounding errors.

\begin{table}
\caption{Model Comparison on Grayscale Set11 Dataset, Using Averaged PSNR (dB) and SSIM Metrics (1st  \& 2nd Column under each Sampling Rate)}
\label{table: compare set 11}
\setlength{\tabcolsep}{3 pt}
\centering
\begin{tabular}{|c|c|cc|cc|cc|}
\hline
Dataset&  Model&  \multicolumn{6}{c|}{Sampling Rate}  \\
\cline{3-8}
 & & \multicolumn{2}{c}{0.1} &  \multicolumn{2}{c}{0.25} & \multicolumn{2}{c|}{0.3} \\
\hline
\multirow{12}{*}{Set11}  & DAMP \cite{7457256} & 19.87 & 0.376 & 31.62 & 0.723 & 32.64 & 0.754  \\
 & ReconNet \cite{8379450} & 27.63 & 0.849 & 32.07 & 0.925 & 33.17 & 0.938 \\
 & TVAL3 \cite{RePEc:spr:coopap:v:56:y:2013:i:3:p:507-530} & 22.45 & 0.376 & 27.63 & 0.624 & 29.00 & 0.676 \\
 & CSNet\textsuperscript{+} \cite{8765626} & 27.76 & 0.851 & 32.76 & 0.932 & 33.90 & 0.945 \\
 & LDIT \cite{10.5555/3294771.3294940} & 25.56 & 0.769 & 31.35 & 0.904 & 32.69 & 0.922 \\
 & LDAMP \cite{10.5555/3294771.3294940} & 24.94 & 0.748 & 29.93 & 0.878 & 32.01 & 0.914 \\
 & DPDNN \cite{8481558} & 24.53 & 0.739 & 30.63 & 0.892 & 32.06 & 0.915 \\
 & NN \cite{8878159} & 22.99 & 0.659 & 26.57 & 0.784 & 27.64 & 0.810 \\
 & ISTA-NET\textsuperscript{+}  \cite{8578294} & 25.93 & 0.784 & 32.27 & 0.917 & 33.66 & 0.933 \\
 & BCS-Net \cite{9159912} & 29.43 & 0.868 & - & - & 35.60 & 0.955 \\
 & AMP-NET \cite{9298950} & 29.45 & 0.879 & \textbf{34.59} & 0.948 & \textbf{35.90} & 0.957 \\
 & R\textsuperscript{2}CS-NET\textsuperscript{1ch}* & \textbf{29.88} & \textbf{0.936} & 34.29 & \textbf{0.974}  & 35.14 & \textbf{0.978} \\
\hline
\multicolumn{8}{l}{* This is a set of single-channel models trained for grayscale image CS.} \\
\end{tabular}
\end{table}

\section{Conclusions and Future Work}
This work proposes a novel deep CS framework, termed as R\textsuperscript{2}CS-NET. Inspired by iterative online optimizers, a recurrent-residual structural constraint is explored for image CS. As an alternative for unrolled NN, it possess improved robustness and generalization capacity while maintaining high efficiency. The R\textsuperscript{2}CS-NET demonstrates leading CS quality and measurement noise robustness among existing benchmarks. Designs inspired by more advanced optimization models can be further investigated in specific application domains.

\bibliographystyle{IEEEtran}
\bibliography{IEEEabrv,deep_cs}

\end{document}